*To be submitted to Phys. Rev. B*

# Resolution Function Theory in Piezoresponse Force Microscopy: Domain Wall Profile, Spatial Resolution, and Tip Calibration

Anna N. Morozovska* and Svetlana L. Bravina

Institute of Physics, National Academy of Science of Ukraine,

46, pr. Nauki, 03028 Kiev, Ukraine

Eugene A. Eliseev

Institute for Problems of Materials Science, National Academy of Science of Ukraine,

3, Krjijanovskogo, 03142 Kiev, Ukraine

Sergei V. Kalinin†

Materials Science and Technology Division, Oak Ridge National Laboratory

Oak Ridge, TN 37831

Piezoresponse Force Microscopy (PFM) has emerged as a primary tool for imaging, domain engineering, and switching spectroscopy on ferroelectric materials. Quantitative interpretation of PFM data including measurements of the intrinsic width of the domain walls, geometric parameters of the domain below the tip in local hysteresis loop measurements, as well as interpretation of switching and coercive biases in terms of materials properties and switching mechanisms, requires reliable knowledge on electrostatic field structure produced by the tip. Using linear imaging theory, we develop a theoretical approach for interpretation of these measurements and determination of tip parameters from a calibration standard. The resolution and object transfer functions in PFM are derived and effect of materials parameters on resolution is determined. Closed form solutions for domain wall profiles in vertical and lateral PFM and signal from cylindrical domain in transversally isotropic piezoelectric are derived for point-charge and sphere-plane geometry of the tip.

* Permanent address: V.Lashkaryov Institute of Semiconductor Physics, National Academy of Science of Ukraine, 41, pr. Nauki, 03028 Kiev, Ukraine

† Corresponding author, sergei2@ornl.gov

## I. Introduction

Ferroelectric materials have found numerous applications as ferroelectric non-volatile memories,[1,2] sensors and actuators,[3,4] and are developed as the next generation data storage media.[5] The polarization-dependent chemical reactivity of ferroelectric surfaces in acid dissolution[6] and metal photodeposition[7] processes is suggested as a basis for novel nanofabrication methods.[8,9] In parallel, much interest has recently been attracted to the physics of nanoscale ferroelectrics, including ferroelectricity in epitaxial thin films,[10,11] static and dynamic properties of polar nanoregions in relaxor ferroelectrics,[12] interplay between ferroelectricity and magnetism in multiferroic materials[13,14] and self-assembled nanostructures,[15] novel polar orderings in low-dimensional ferroelectrics,[16] and surface ferroelectricity and piezoelectricity.[17] Piezoresponse Force Microscopy (PFM) based on direct detection of tip-bias induced electromechanical surface displacement has been developed to probe piezoelectric and ferroelectric properties on the nanoscale.[18,19,20,21,22] In particular, detection of the amplitude and phase of surface displacement due to the inverse piezoelectric effect allows local electromechanical activity to be mapped, from which domain structure and orientation, presence of domain walls, etc. can be inferred with ~10 nm resolution. More generally, electromechanical activity in polar materials is directly related to the lattice instabilities and phonon mode softening, thus providing information on physics of these materials on the nanoscale. Particularly of interest for ferroelectric materials are local PFM hysteresis measurements (Piezoresponse Force Spectroscopy),[23,24] in which the dc-bias dependence of electromechanical signal contains the information of the size of nascent domain below the tip[25] and hence on the kinetics and thermodynamics of tip-induced switching process.[26] Furthermore, the type of switching, bias- or strain-induced phase transitions, etc. can be probed.

In the last five years, advances in scanning probe microscopy (SPM) and sample preparation routines have extended applicability of PFM to a broad range of piezoelectric materials. Imaging of III-V nitrides[27] and biopolymers[28,29] in calcified and connective tissues has successfully been demonstrated. PFM provides an approach to study local structure, including aspects inaccessible by other techniques, i.e., Ga- and N- surface termination and inversion boundaries in piezoelectric III-V nitrides, collagen orientation and molecular ordering in calcified tissues, etc. Demonstrated sub-10 nm resolution and the smallness of topographic cross-talk that plagues most property-sensitive SPM technique suggest immense potential of PFM for characterization of polar materials and biological systems.

The unique feature of PFM that differentiates it from traditional force- or current-based SPM is that the PFM signal is only weakly dependent on the apparent contact area[30,31] and is determined only by the tip-induced potential on the surface. This limits applicability of conventional resonance enhancement to amplify the magnitude of weak (~10-100 pm) surface displacement, since local contact resonant frequency is primarily determined by surface topography and elasticity,[32] and imposes stringent requirements on the noise level and detector sensitivity. However, this unique aspect of bias-dependent contact mechanics also renders low-frequency PFM intrinsically quantitative and relatively insensitive to topographic cross-talk. Combined with the possibility of detection of all three components of electromechanical response vector, this opens the pathway to molecular orientation imaging on the nanoscale.[33]

To date, multiple applications of PFM for imaging ferroelectric and piezoelectric materials and spectroscopy of polarization switching processes in ferroelectrics have been established.[34,35,36] These applications necessitate the development of quantitative imaging theory for PFM in order to

- establish the resolution and information limits in PFM and its dependence on tip geometry and materials properties, suggesting strategies for high-resolution imaging;
- develop the pathways for calibration of tip geometry in PFM experiment for quantitative data interpretation;
- interpret the imaging and spectroscopy data in terms of intrinsic domain wall widths and the size of nascent domain below the tip; and
- reconstruct the ideal image from experimental data, effectively deconvolution tip contribution, and establish applicability limits and errors associated with such deconvolution process.

The phenomenological resolution function theory for PFM was recently developed by Kalinin et al.[37] Here, we derive analytical forms for resolution and object transfer functions in PFM using Green's function based decoupled theory. The basic principles of linear imaging theory, the definitions of resolution and information limit and their relationship to PFM amplitude and phase signals are discussed in Section II. The analytical form of the resolution function in decoupled Green's function approach is derived in Section III. The analytical expressions for domain wall profiles for various tip geometries are derived, and resolution limits anticipated for individual domains and periodic domain structures in ferroelectric materials are compared in Section IV. PFM signal for cylindrical domains and nested cylindrical domains, providing an approach for interpretation of data in PFM spectroscopy, is analyzed in Section V. Information limit in PFM, i.e. the size of minimal domain that can still be reliably visualized, is discussed in Section VI. Further perspectives on quantitative PFM probing are given in Section VII.

## II. Principles of PFM and Linear Imaging Theory

In PFM, a conductive tip, biased with $V_{tip} = V_{dc} + V_{ac}\cos(\omega t)$, is brought into contact with the surface, and the electromechanical response of the surface is detected as the first harmonic component of bias-induced tip deflection, $p = p_0 + p_{1\omega}\cos(\omega t + \varphi)$.[34,35] The phase of the response, $\varphi$, yields information on the polarization direction below the tip. For $c^-$ domains (polarization vector pointing downward) the application of a positive tip bias results in the expansion of the sample and surface oscillations are in phase with the tip voltage, $\varphi = 0$. For $c^+$ domains $\varphi = 180°$. Traditionally, the PFM signal is plotted either as a pair of amplitude-phase, $A = p_{1\omega}/V_{ac}$, $\varphi$, images, or a mixed signal representation in which the piezoresponse, $PR = A\cos\varphi$, is used.

The resolution and probed volume in PFM is determined by the structure of electroelastic fields inside the material, specifically the voltage derivative of the normal displacement field, $\partial u_3(\mathbf{x})/\partial V$.[30,31] In general, calculation of the electroelastic fields in the material requires the solution to a coupled problem, which is currently available only for a transversally isotropic case and is also limited to the electric field produced in the contact area. A simplified approach suggested by Felten et al.,[38] and Scrymgeour and Gopalan[39] is based on the solution to a decoupled problem. In this case, the electric field in the material is calculated using a rigid electrostatic model (no piezoelectric coupling); the strain or stress field is calculated using constitutive relations for a piezoelectric solid, and the displacement field is evaluated using an appropriate Green's function for an isotropic or anisotropic solid. In this approximation, PFM signal, i.e., surface displacement $u_i(\mathbf{x},\mathbf{y})$ at location $\mathbf{x}$ induced by the tip at position $\mathbf{y}=(y_1,y_2)$ is given by

$$u_i(\mathbf{x},\mathbf{y})=\int_{-\infty}^{\infty}d\xi_1\int_{-\infty}^{\infty}d\xi_2\int_{0}^{\infty}d\xi_3\frac{\partial G_{ij}(x_1-\xi_1,x_2-\xi_2,\xi_3)}{\partial\xi_k}E_l(\xi)c_{kjmn}d_{lnm}(y_1+\xi_1,y_2+\xi_2,\xi_3) \qquad (1)$$

Here coordinate $\mathbf{x}=(x_1,x_2,z)$ is linked to the indentor apex, coordinates $\mathbf{y}=(y_1,y_2)$ denote indentor position in the sample coordinate system $\mathbf{y}$ (Fig. 1). Coefficients $d_{mnk}$ and $c_{jlmn}$ are position dependent components of the piezoelectric strain constant and elastic stiffness tensors, respectively. $E_k(\mathbf{x})$ is the electric field strength distribution produced by the probe. The Green's function for a semi-infinite medium $G_{3j}(\mathbf{x}-\xi)$ links the eigenstrains $c_{jlmn}\ d_{mnk}E_k$ to the displacement field.

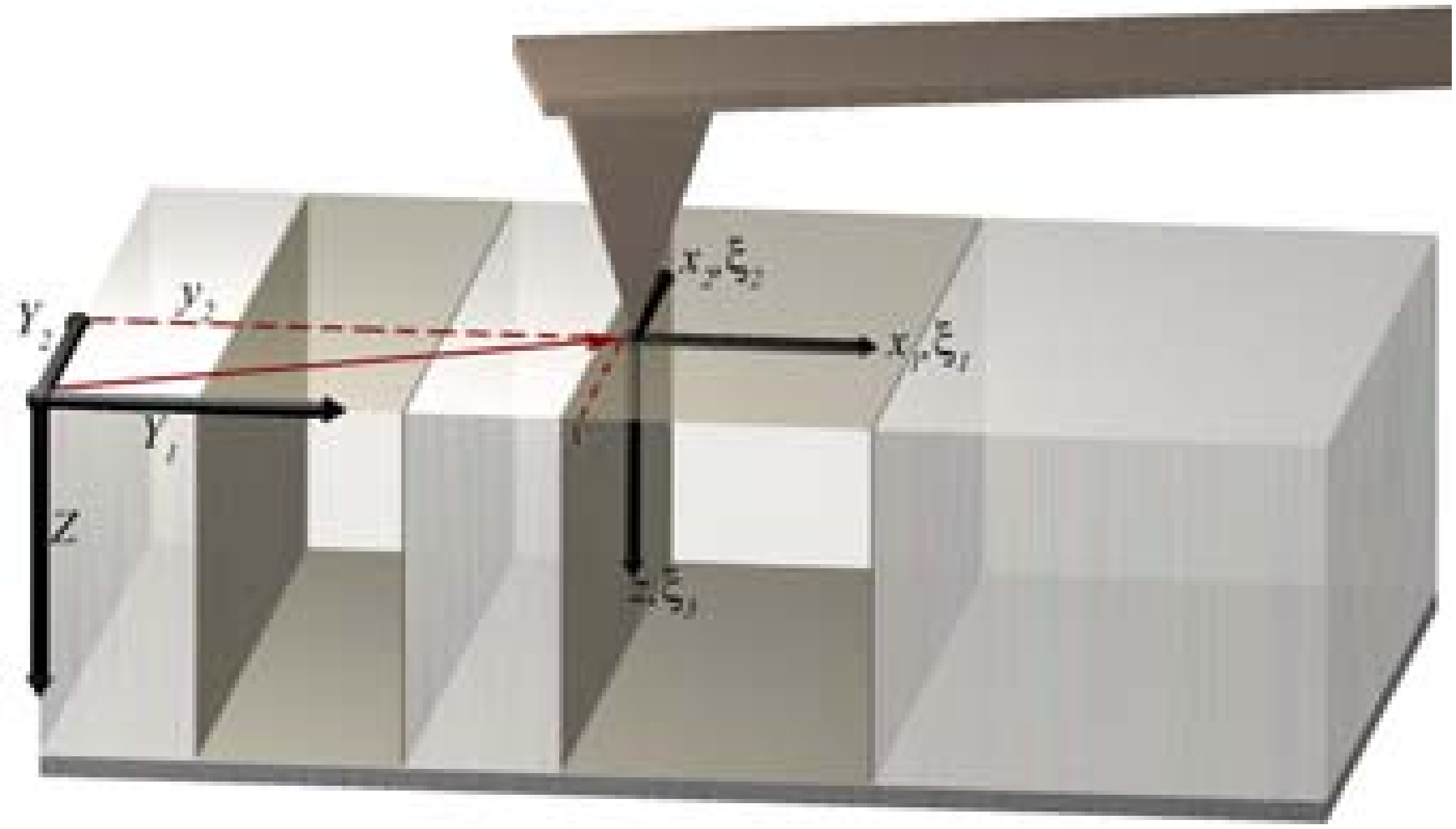

FIG. 1. (Color online) Coordinate systems in PFM experiment.

If the sample is uniform in $z$-direction on the scale of the penetration depth of electric field, i.e., $c_{jlmn}d_{mnk}(\mathbf{x},z)\approx c_{jlmn}d_{mnk}(\mathbf{x})$, vertical surface displacement below the tip, i.e., vertical PFM signal, can be rewritten as

$$u_3(\mathbf{0},\mathbf{y})=\int_{-\infty}^{\infty} d_{mnk}(\mathbf{y}-\boldsymbol{\xi})\left(\int_{z=0}^{\infty} c_{jlmn}E_k(-\xi_1,-\xi_2,z)\frac{\partial}{\partial\xi_l}G_{3j}(\xi_1,\xi_2,z)dz\right)d\xi_1 d\xi_2, \qquad (2)$$

i.e., as a convolution of a function describing the spatial distribution of material properties, $d_{mnk}(\mathbf{x})$, and a function related to probe parameters (integral in parenthesis).

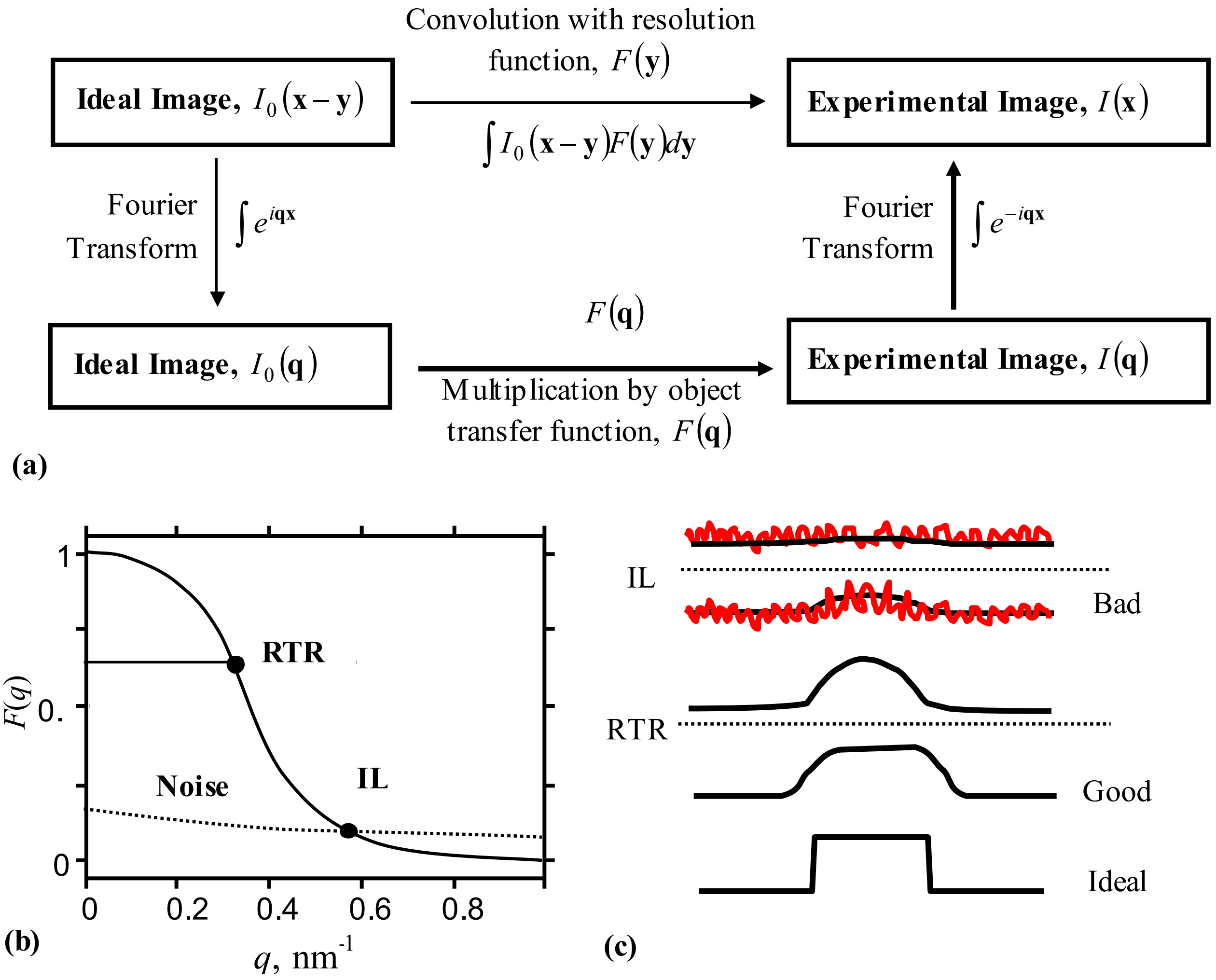

FIG. 2. (Color online). Linear Imaging theory. (a) Relation between ideal image and experimental image in real and Fourier spaces. Object transfer function is a Fourier transform of resolution function. (b) Definitions of Rayleigh two-point resolution (RTR) and information limit (IL). (c). Domain signal for different resolutions. For domain sizes larger then RTR, material properties can be determined quantitatively. Below RTR, but above IL, the presence of domain can be established, but properties could not be measured. Below IL, the signal is below noise level and domain presence could not be established.

Image formation described by Eq. (2) belongs to the class of so-called linear imaging mechanisms. In scalar case, the measured image $I(\mathbf{x})$, where $\mathbf{x}$ is a set of spatial coordinates, is given by the convolution of an *ideal image* (representing material properties) $I_0(\mathbf{x}-\mathbf{y})$ with the resolution function dependent on probe geometry, $F(\mathbf{y})$:

$$I(\mathbf{x}) = \int I_0(\mathbf{x}-\mathbf{y})F(\mathbf{y})d\mathbf{y} + N(\mathbf{x}) \tag{3}$$

where $N(\mathbf{x})$ is the noise function. The Fourier transform of Eq. (3) is

$$I(\mathbf{q}) = I_0(\mathbf{q})F(\mathbf{q}) + N(\mathbf{q}) \tag{4}$$

where $I(\mathbf{q}) = \int I(\mathbf{x})e^{i\mathbf{qx}}d\mathbf{x}$, $I_0(\mathbf{q})$, and $N(\mathbf{q})$ are the Fourier transforms of the measured image, ideal image, and noise, respectively. The object transfer function (OTF), $F(\mathbf{q})$, is a Fourier transform of the resolution function, $F(\mathbf{y})$, and describes the transfer between Fourier components of ideal and experimental image.

### II.1. Resolution and Information Limit

In the framework of linear imaging theory, the Rayleigh two-point resolution (RTR) of the technique can be defined as a full width at half maximum of the object transfer function, $F(\mathbf{q})$, as $1/\mathbf{q}$ for which $F(\mathbf{q})/F(0) = 0.5$ (Fig. 2).[37] RTR defines the minimal size of the uniform object required for quantitative property measurements. As applied to ferroelectric materials, RTR will determine e.g., the minimum domain size for which PFM signal in the center of the domain achieves saturated value for infinite material.

The second fundamental characteristic of resolution is information limit (IL), which describes the minimal size of the feature (e.g., domain) that can be reliably detected. The characteristic feature of PFM, as well as many other property-sensitive SPM techniques such as Magnetic Force Microscopy and Kelvin-Probe Force Microscopy is that the resolution function does not have zeroes imposed by the physics of imaging process. Hence the minimal feature size that can be detected is limited by the thermal noise level of the systems (dynamic disorder) and topographic cross-talk (frozen disorder). In particular, in PFM the information limit is determined by condition $F(\mathbf{q})/F(0) > N(\mathbf{q})$.

Finally, in PFM both mixed-signal and phase images can be collected. It has been argued by Kalinin et al., that RTR of phase image is similar to the IL of the mixed signal image.[37] Since the phase signal does not contain any information other then the sign (upward or downward) of predominant polarization orientation, here we consider the resolution theory for mixed signal only.

### II.2. Phenomenological determination of resolution function

The object transfer function, $F(\mathbf{q})$, and the resolution function, $F(\mathbf{y})$, can be determined directly from experimental image, $I(\mathbf{q})$, provided that the ideal image, $I_0(\mathbf{q})$, is known. Once the resolution function is determined for a known calibration standard, it can be used to extract the ideal image, $I_0(\mathbf{x})$, from a measured image, $I(\mathbf{x})$ for an arbitrary sample. This approach for PFM using model $Pb(Zr,Ti)O_3$ (PZT) system has been demonstrated.[37] However, the resolution function $F(\mathbf{y})$ in PFM as defined by Eq. (2) depends on the electrostatic field generated by the tip. As such, it also depends on dielectric properties of material and hence phenomenological resolution function for one material could not be used for image reconstruction in a different material, necessitating development of more general theory as described below.

## II.3. Resolution function vs. tip geometry

Analytically, the resolution and transfer functions are determined by the tip geometry, dielectric properties of materials and surrounding media, and elastic properties of material, that determine electric field distribution and local elastic Green's function. While materials properties vary from point to point, the tip geometry can be assumed to be material-independent.[40] Hence, calculation of resolution function, or, the reverse problem, calibration of tip parameters from a known standard, should ultimately relate tip geometry and resolution function. For material with inhomogeneous piezoelectric, elastic, and dielectric properties, the resolution function is position dependent and the PFM reconstruction problem is extremely difficult (albeit possible if tip geometry is known).

Considered here is a special, and particularly important, class of problems related to 180° domain walls perpendicular to the surface, often encountered in the materials systems used for data storage and ferroelectric recording media. In this case, the dielectric and mechanical properties defined by rank 2 and rank 4 tensors respectively do not change across the domain wall. At the same time, piezoelectric properties change sign across the domain wall. Hence, the resolution function in Eq. (2) is constant across the sample and position independent, and the problem of tip calibration is reduced to determination of position-independent electric field structure produced by the tip. Below, we derive analytical expressions for resolution function, and analyze the domain wall profile in vertical and lateral PFM and signal from small cylindrical domain.

## III. Resolution Function in PFM

In a PFM experiment, the surface is contacted by the biased tip (also referred to as indentor). Here, we generalize Eq. (2) to give surface displacement $u_i(\mathbf{x},\mathbf{y})$ at location $\mathbf{y}$ induced by the tip at position $\mathbf{x}$:

$$u_i(\mathbf{x},\mathbf{y}) = \int_{-\infty}^{\infty} d\xi_1 \int_{-\infty}^{\infty} d\xi_2 \int_{0}^{\infty} d\xi_3 \frac{\partial G_{ij}(x_1-\xi_1, x_2-\xi_2, \xi_3)}{\partial \xi_k} E_l(\xi)\cdot c_{kjmn} d_{lnm}(y_1+\xi_1, y_2+\xi_2, \xi_3) \quad (5)$$

The electric field $E_k(\mathbf{x}) = -\partial V_Q / \partial x_k$ is produced by the tip in the point $\mathbf{x} = (x_1, x_2, z)$ on the sample, $d_{klj}(\mathbf{y})$ are the stress piezoelectric tensor components representing material properties (*ideal image*), $c_{kjmn} = c_{kjmn}(\mathbf{y})$ are stiffness tensor components. Coordinate systems $\mathbf{x}$ and $\xi$ are linked to the indentor, coordinates $\mathbf{y} = (y_1, y_2, z)$ is indentor apex position in the sample coordinate system $\mathbf{y}$ (see Fig.1).

Eq. (5) contains a large number of tensor components of the elastic Green's function and position-dependent electric field determined by dielectric properties of material and tip geometry. Below, we discuss the approximations involved in Green's function choice and the electric field structure in material.

### III.1. Elastic Green's function

For most inorganic ferroelectrics, the elastic properties of material are weakly dependent on orientation. Hence, material can be approximated as elastically isotropic. Corresponding Green's tensor for elastic isotropic half-plane is given by Lurie[41] and Landau and Lifshitz[42]:

$$
G_{ij}(x_1,x_2,\xi_3)=\begin{cases}
\dfrac{1+\nu}{2\pi Y}\left[\dfrac{\delta_{ij}}{R}+\dfrac{(x_i-\xi_i)(x_j-\xi_j)}{R^3}+\dfrac{1-2\nu}{R+\xi_3}\left(\delta_{ij}-\dfrac{(x_i-\xi_i)(x_j-\xi_j)}{R(R+\xi_3)}\right)\right] & i,j\neq 3\\
\dfrac{(1+\nu)(x_i-\xi_i)}{2\pi Y}\left(\dfrac{-\xi_3}{R^3}-\dfrac{(1-2\nu)}{R(R+\xi_3)}\right) & i=1,2 \quad and \quad j=3\\
\dfrac{(1+\nu)(x_j-\xi_j)}{2\pi Y}\left(\dfrac{-\xi_3}{R^3}+\dfrac{(1-2\nu)}{R(R+\xi_3)}\right) & j=1,2 \quad and \quad i=3\\
\dfrac{1+\nu}{2\pi Y}\left(\dfrac{2(1-\nu)}{R}+\dfrac{\xi_3{}^2}{R^3}\right) & i=j=3
\end{cases}
\tag{6}
$$

where $R=\sqrt{(x_1-\xi_1)^2+(x_2-\xi_2)^2+\xi_3^2}$, $Y$ is Young's modulus, and $\nu$ is the Poisson ratio. Stiffness tensor $c_{kjmn}$ corresponds to the elastically isotropic medium $c_{klmn}=\dfrac{Y}{2(1+\nu)}\left[\dfrac{2\nu}{1-2\nu}\delta_{kl}\delta_{mn}+\delta_{km}\delta_{ln}+\delta_{kn}\delta_{lm}\right]$.

Here we specifically address the choice of ideal image and Green's function representations. In the early version of the theory,[37,43] the ideal PFM image was identified with the distribution of piezoelectric stress coefficient, $e_{ijk}$. However, sensitivity analysis of the exact solution for the transversally isotropic homogeneous system[30] has shown that piezoelectric response actually depends primarily on piezoelectric strain components $d_{ijk}$, and dielectric anisotropy factor, $\gamma=\sqrt{\varepsilon_{33}/\varepsilon_{11}}$. In this ($c_{kjmn}$, $d_{ijk}$, $\varepsilon_{ij}$) representation piezoresponse is virtually independent on elastic properties. At the same time, in ($c_{kjmn}$, $e_{ijk}$, $\varepsilon_{ij}$) representation piezoresponse strongly depends on elastic properties. Hence, here we derive theory utilizing the piezoelectric strain elements. In addition to sensitivity analysis, in this case the result of convolution $G_{ij,k}c_{kjmn}$ does not depend on the Young's modulus, and elastic part of the resolution function depends only on the Poisson ratio of material, which varies only weakly between dissimilar materials.

### III.2. Electrostatic field structure

The key component of PFM signal is the electric field distribution in the material controlled by the geometric parameters of the tip and dielectric properties of material and medium. The detailed analysis of electric field structure for material of arbitrary symmetry is given elsewhere.[43] In particular, for isotropic and transversally isotropic materials symmetries the electrostatic problems for spherical tip geometry can be solved using image charge method, in which tip is represented by the set of image charges chosen so that corresponding isopotential contour represents tip geometry. Similar image charge methods provide good approximation for lower material symmetries and other tip geometries (e.g., line charge model for conical part of the tip).

For the case of dielectrically transverse isotropic ferroelectric the potential $V_Q$ in the linear point charges models of the tip has the form:

$$V(\rho, z) = \frac{1}{2\pi\varepsilon_0(\varepsilon_e + \kappa)} \sum_{m=0}^{\infty} \frac{Q_m}{\sqrt{\rho^2 + (z/\gamma + d_m)^2}}., \qquad (7)$$

where $\sqrt{x_1^2 + x_2^2} = \rho$ and $\xi_3 = z$ are the radial and vertical coordinates respectively, $\varepsilon_e$ is the dielectric constant of the ambient, $\kappa = \sqrt{\varepsilon_{33}\varepsilon_{11}}$ is effective dielectric constant of material, $\gamma = \sqrt{\varepsilon_{33}/\varepsilon_{11}}$ is the dielectric anisotropy factor, $-d_m$ is the z-coordinates of the point charge $Q_m$ and summation is performed over the set of image charges representing the tip.

The potential in the sphere-plane model can be obtained from Eq. (7), where the summation is performed over image charges. In the case of rigorous sphere-plane model of the tip of curvature $R_0$ located at distance $\Delta R$ from the sample surface, the image charges are given by recurrent relations $d_{m+1} = R_0 + \Delta R - \frac{R_0^2}{R_0 + \Delta R + d_m}$ and $Q_{m+1} = \frac{\kappa - \varepsilon_e}{\kappa + \varepsilon_e} \cdot \frac{R_0}{R_0 + \Delta R + d_m} Q_m$, where $Q_0 = 4\pi\varepsilon_0\varepsilon_e R_0 U$, $d_0 = R_0 + \Delta R$ and $U$ is tip bias (see e.g., Ref. 30).

The rigorous sphere-plane model involves summation over large number of image charges. An alternative approach to describe electric fields in the immediate vicinity of the tip-surface junction is the use of effective point charge model, in which the charge magnitude and charge-surface separation are selected such that corresponding isopotential surface reproduces tip radius of curvature and tip potential. In this case, the tip is represented by a single charge $Q = 2\pi\varepsilon_0\varepsilon_e R_0 U(\kappa + \varepsilon_e)/\kappa$ located at $d = \varepsilon_e R_0/\kappa$.[25]

### III.3 Resolution function of the point charge

In the case when $x_1 = x_2 = 0$ (response below the tip) and strain piezoelectric coefficient $d_{klj}(\xi)$ is independent on $\xi_3$ (system is uniform in *z*-direction), the measured piezoresponse $u_i(0,\mathbf{y}) \equiv u_i(\mathbf{y})$ is given by the convolution of an *ideal image* $d_{klj}(\mathbf{y} - \xi)$ with the resolution function $W_{ijkl}(\xi)$:

$$u_i(\mathbf{y}) = \int_{-\infty}^{\infty} d\xi_1 \int_{-\infty}^{\infty} d\xi_2 \, W_{ijkl}(-\xi_1, -\xi_2) d_{lkj}(y_1 - \xi_1, y_2 - \xi_2) \qquad (8a)$$

where

$$W_{ijkl}(\xi_1,\xi_2) = c_{kjmn}\int_0^\infty d\xi_3 \frac{\partial G_{im}(-\xi_1,-\xi_2,\xi_3)}{\partial \xi_n} E_l(\xi_1,\xi_2,\xi_3) \qquad (8b)$$

In general case, the resolution function components can be calculated numerically (Pade approximations for several components are given in Appendix A). For the transversally isotropic dielectric media only three components $W_{333}$, $W_{313}$ and $W_{351}$ contribute into vertical displacement $u_3$ (in Voigt representation). In most cases, the component $W_{333}$ corresponding to piezoelectric constant $d_{33}$ provides the dominant (> 50%) contribution to the response.[44]

Taking into account the linearity of the theory, the resolution function (4) could be summarized over the series of the image charges $Q_m$ located at distances $d_m$ within the framework of the sphere-plane model of the tip. Hence, here we derive the resolution function from the point charge, from which resolution functions for other charge distributions can be obtained by summation or integration over image charges.

The Fourier transform of Eq. (8a) is

$$\tilde{u}_i(\mathbf{q}) = \tilde{d}_{klj}(\mathbf{q})\tilde{W}_{ijkl}(-\mathbf{q}) + N(\mathbf{q}) \qquad (9)$$

where $\tilde{u}_i(\mathbf{q}) = \int u_i(\mathbf{x})e^{i\mathbf{qx}}d\mathbf{x}$, $\tilde{d}_{klj}(\mathbf{q})$ and $N(\mathbf{q})$ are the Fourier transforms of the measured image, ideal image, and noise respectively. The tensorial OTF $\tilde{W}_{ijkl}(\mathbf{q})$ is defined as a Fourier transform of the resolution function $\tilde{W}_{ijkl}(\mathbf{q})$, namely:

$$\tilde{W}_{ijkl}(\mathbf{q}) = c_{kjmn}\int_{-\infty}^{\infty} dk_1 \int_{-\infty}^{\infty} dk_2 \int_0^\infty d\xi_3 \frac{\partial \tilde{G}_{im}(q_1-k_1,q_2-k_2,\xi_3)}{\partial \xi_n} \tilde{E}_l(k_1,k_2,\xi_3) \qquad (10)$$

In general case the resolution function components should be calculated numerically. For the dielectrically transverse isotropic ferroelectric (3) $\tilde{W}_{ijk}(\mathbf{q})$ depends only on the absolute value of wave vector, $q$ (see Appendix A). Using the notations of Eq. (4), the rotationally invariant object transfer function can be derived as

$$F_{3q}(q) = \widetilde{W}_{333}(q)d_{33} + \widetilde{W}_{313}(q)d_{31} + \widetilde{W}_{351}(q)d_{15}, \tag{11a}$$

where the ideal image now represents the domain structure (i.e., the elements of the piezoelectric constant tensor). Note that the full tensorial theory Eqs. (9, 10) is applicable for arbitrary distributions of $d_{klj}(\mathbf{x})$ provided that elastic and dielectric properties do not change. The scalar theory Eqs. (3, 4) is applicable if domain structure can be defined by single parameter (up-down orientation) or, more generally, if all elements of piezoelectric constant tensor scale proportionately, $d_{klj}(\mathbf{x}) \sim \alpha(\mathbf{x})d_{klj}$. Finally, Eq. (11a) is valid for transversally isotropic materials in which $d_{klj}(\mathbf{x}) \sim \alpha(\mathbf{x})d_{klj}$, and thus represents the most broadly encountered case.

After lengthy integration and Pade analysis, the approximate analytical expressions for the non-zero components in Eq. (11a) are derived as:

$$\widetilde{W}_{333}(q) \approx -\frac{Q}{2\pi\varepsilon_0(\kappa + \varepsilon_e)d}\left(\frac{\gamma q d}{2} + \frac{(1+\gamma)^2}{1+2\gamma}\right)^{-1}, \tag{11b}$$

$$\widetilde{W}_{313}(q) \approx -\frac{Q}{2\pi\varepsilon_0(\kappa + \varepsilon_e)d}\left(-\left(\gamma q d + \frac{(1+\gamma)^2}{\gamma}\right)^{-1} + (1+2\nu)(\gamma q d + 1 + \gamma)^{-1}\right), \tag{11c}$$

$$\widetilde{W}_{351}(q) \approx -\frac{Q}{2\pi\varepsilon_0(\kappa + \varepsilon_e)d}\left(\frac{\gamma(qd)^3}{6} + 2(qd)^2 + \frac{16-15\gamma^2}{4\gamma}qd + \frac{(1+\gamma)^2}{\gamma^2}\right)^{-1}, \tag{11d}$$

where $\sqrt{q_1^2 + q_2^2} = q$. The dependence of components $\widetilde{W}_{3ij}$ on wavevector $q$ is shown in Fig. 3. Note, that quasi-Lorenz dependences in Eq. (11b-d), as anticipated from $1/r$ behavior of Green's function for point force/charge on piezoelectric surface.[45] Here, approximation

Eq. (11d) is valid for $\gamma < 1$, whereas for $\gamma \geq 1$ the linear term $\sim qd$ that caused unphysical pole should be omitted (as the result, the accuracy of Pade approximation decreases).

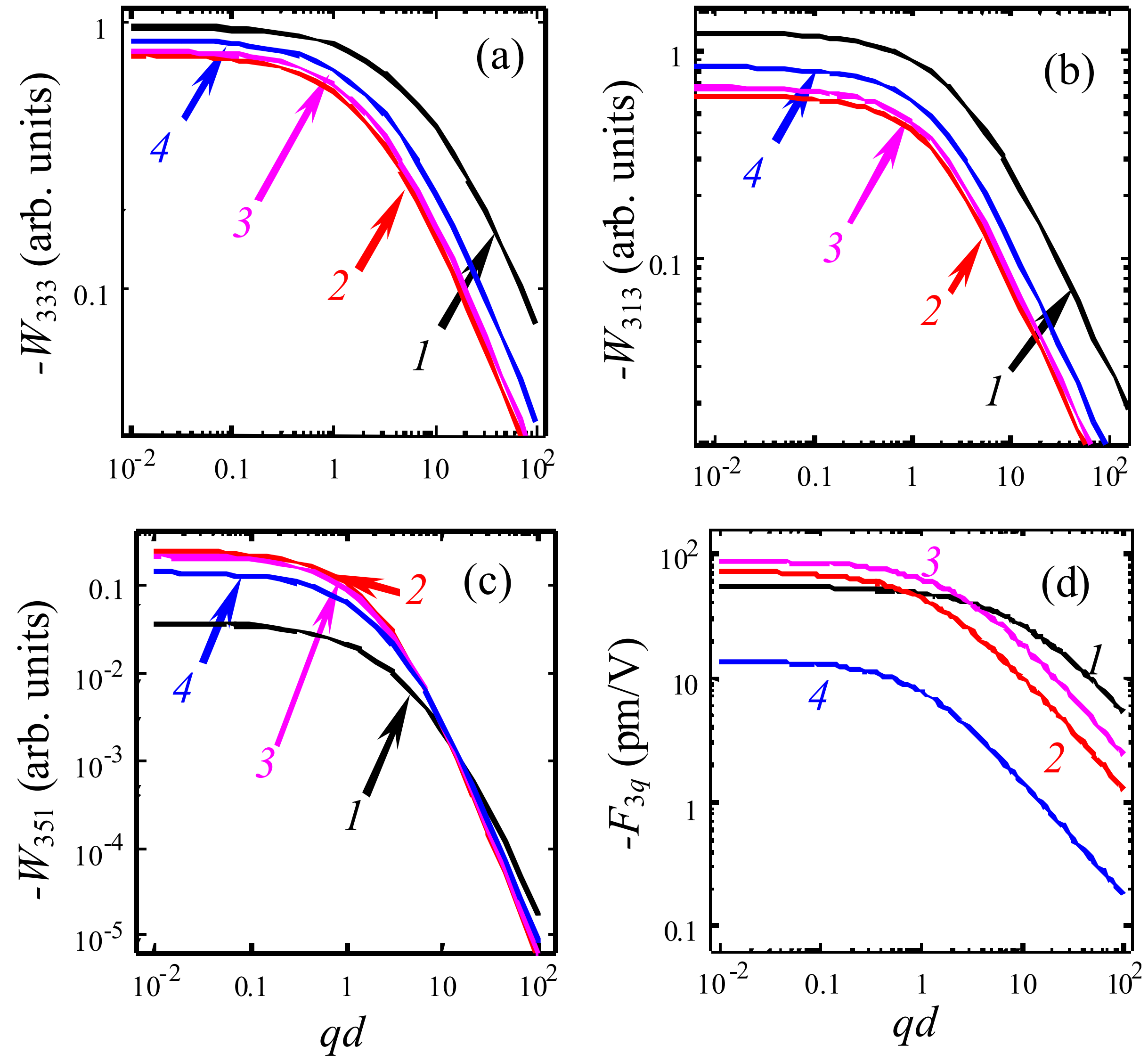

FIG. 3. (Color online). The resolution function components $\widetilde{W}_{3ij}/V_Q(\mathbf{0})$ (a,b,c) and $F_{3q}(q)$ (d) vs. the wave vector absolute value $q$ for different ferroelectric materials $BaTiO_3$ ($\gamma = 0.24$, curve 1), PZT6B ($\gamma = 0.99$, curve 2), $PbTiO_3$ ($\gamma = 0.87$, curve 3), $LiNbO_3$ ($\gamma = 0.60$, curve 4).

Note, that the contribution of non-zero component $d_{22}$ on $LiNbO_3$ piezoresponse is absent in the homogeneous case;[43] its possible role depending on the domain wall orientation with respect to the crystal axes will be discussed elsewhere.

The FWHM of the object transfer function is inversely proportional to charge-surface separation, *d*, and strongly depends on dielectric anisotropy, $\gamma$. For $\tilde{W}_{333}(q)$ component,

$$q_{FWHM} \approx \frac{2}{\gamma d}\frac{(1+\gamma)^2}{1+2\gamma}. \tag{12}$$

Thus, the halfwidth of the original $\rho_{FWHM}$ is proportional to $\gamma d$, where the coefficient of proportionality is independent on *d*, but depends on dielectric anisotropy $\gamma$, as illustrated in Fig. 4. Note that for small dielectric anisotropy, the resolution scales linearly with $\gamma$, while it saturates for $\gamma >> 1$. Hence, $\gamma << 1$ favors high-resolution imaging.

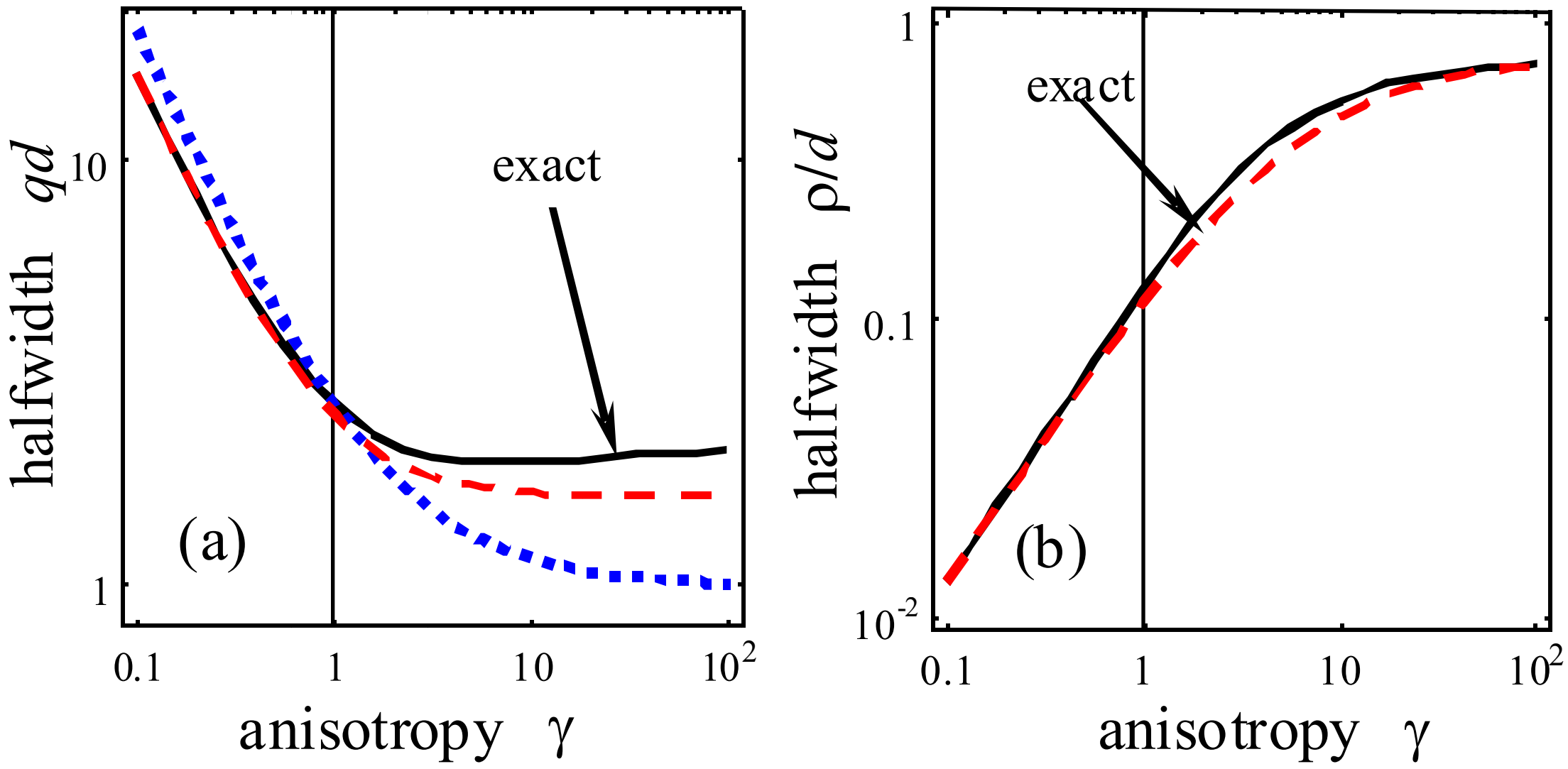

FIG. 4. (Color online). (a) The FWHM for OTF $\tilde{W}_{333}$ vs. the dielectric anisotropy $\gamma$ in Fourier space (RTR) Shown are exact calculations (solid curves) with approximations (A.8) and (12) (dashed and dotted curves correspondingly). (b) FWHM for resolution function in real space. Shown are exact calculation and approximation $\rho_{FWHM} = \gamma\, d/(7.5 + 1.31\gamma)$ (dashed curve). Note that FWHM for OTF and resolution function are not reciprocally related since the corresponding functional forms are non-Gaussian.

The components of the resolution function in real space are significantly more complex (e.g., $W_{333}(\rho)$ is given in Appendix A).

### III.4. Resolution function for the sphere-plane model of the tip

Following the logic in Section III.3, we derive closed-form approximations for the OTF components $\tilde{W}_{ijk}(q)$ and rotationally invariant OTF $F_{3s}(q)$ for commonly used sphere-plane model for the tip. After the summation over image charges, the approximate analytical expressions are derived for the contributors to vertical PFM signal as:

$$F_{3s}(q) = \tilde{W}_{333}(q) d_{33} + \tilde{W}_{313}(q) d_{31} + \tilde{W}_{351}(q) d_{15} \tag{13a}$$

$$\tilde{W}_{333}(q) \approx -U\left(\frac{\varepsilon_e \gamma q R_0}{\varepsilon_e + \kappa} + \frac{(1+\gamma)^2}{1+2\gamma}\right)^{-1} \tag{13b}$$

$$\tilde{W}_{313}(q) \approx -U\left(-\left(2\frac{\varepsilon_e \gamma q R_0}{\varepsilon_e + \kappa} + \frac{(1+\gamma)^2}{\gamma}\right)^{-1} + (1+2\nu)\left(2\frac{\varepsilon_e \gamma q R_0}{\varepsilon_e + \kappa} + 1 + \gamma\right)^{-1}\right) \tag{13c}$$

$$\tilde{W}_{351}(q) \approx -U\left(\frac{2\gamma}{9}\left(\frac{\varepsilon_e}{\kappa + \varepsilon_e} q R_0\right)^3 + \frac{16}{3}\left(\frac{\varepsilon_e}{\kappa + \varepsilon_e} q R_0\right)^2 + \frac{(1+\gamma)^2}{\gamma^2}\right)^{-1} \tag{13d}$$

The dependences of resolution function components $\tilde{W}_{3ij}$ and rotationally invariant $F_3(q)$ on the wave vector $q$ are shown in Fig. 5 for different ferroelectric materials.

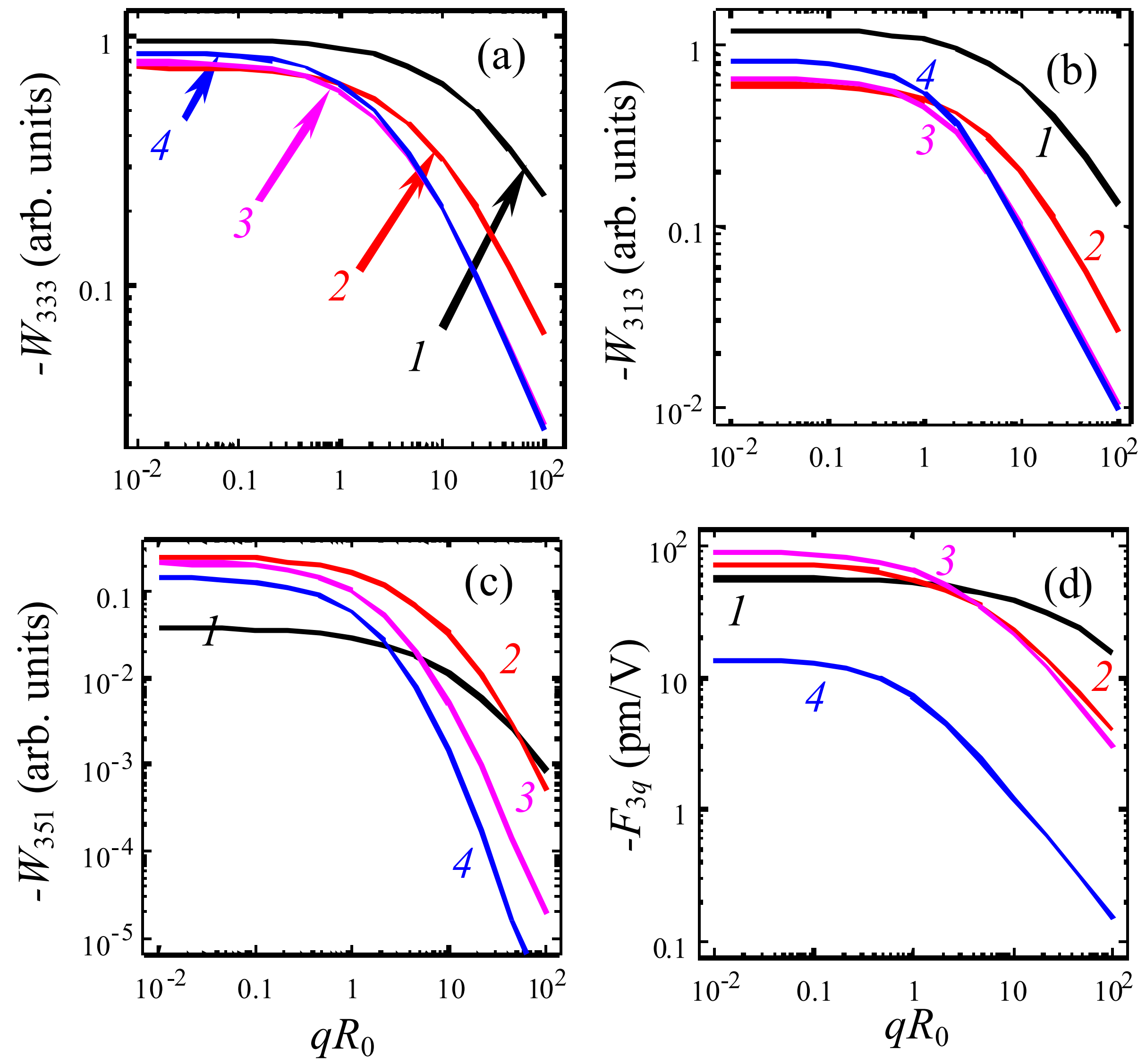

FIG. 5. (Color online). The resolution function components $\widetilde{W}_{3ij}$ (a,b,c) and $F_3(q)$ (d) vs. the absolute value of the wave vector $q$ for different ferroelectric materials $BaTiO_3$ (1), PZT6B (2), $PbTiO_3$ (3), $LiNbO_3$ (4). Calculation was performed within rigorous sphere-plane model of the tip.

From Eq. (13b), the RTR defined from $\widetilde{W}_{333}(q)$ halfwidth is inversely proportional to $R_0$ and depends on $\gamma$ value as

$$q_{FWHM} \approx \frac{\varepsilon_e + \kappa}{\varepsilon_e \gamma R_0} \frac{(1+\gamma)^2}{1+2\gamma}. \qquad (14)$$

The halfwidth of the resolution function $\rho_{FWHM}$ is proportional to $\gamma R_0 \frac{\varepsilon_e}{\varepsilon_e + \kappa}$. The dependence of dimensionless halfwidth $qR_0$ on dielectric anisotropy $\gamma$ and relative permittivity $\kappa/\varepsilon_e$ is illustrated in Fig. 6 for $\widetilde{W}_{333}$. The RTR decreases with dielectric anisotropy factor, $\gamma$, and increases with the dielectric constant of material, $\kappa/\varepsilon_e$. Hence, to increase the lateral resolution $\rho_{FWHM} \sim 1/q_{FWHM}$ in PFM experiments (i.e., in *r*-space) it is desirable to decrease $\gamma$ (imaging of strongly anisotropic materials) and increase $\kappa/\varepsilon_e$ (no water layers and capillary bridges).

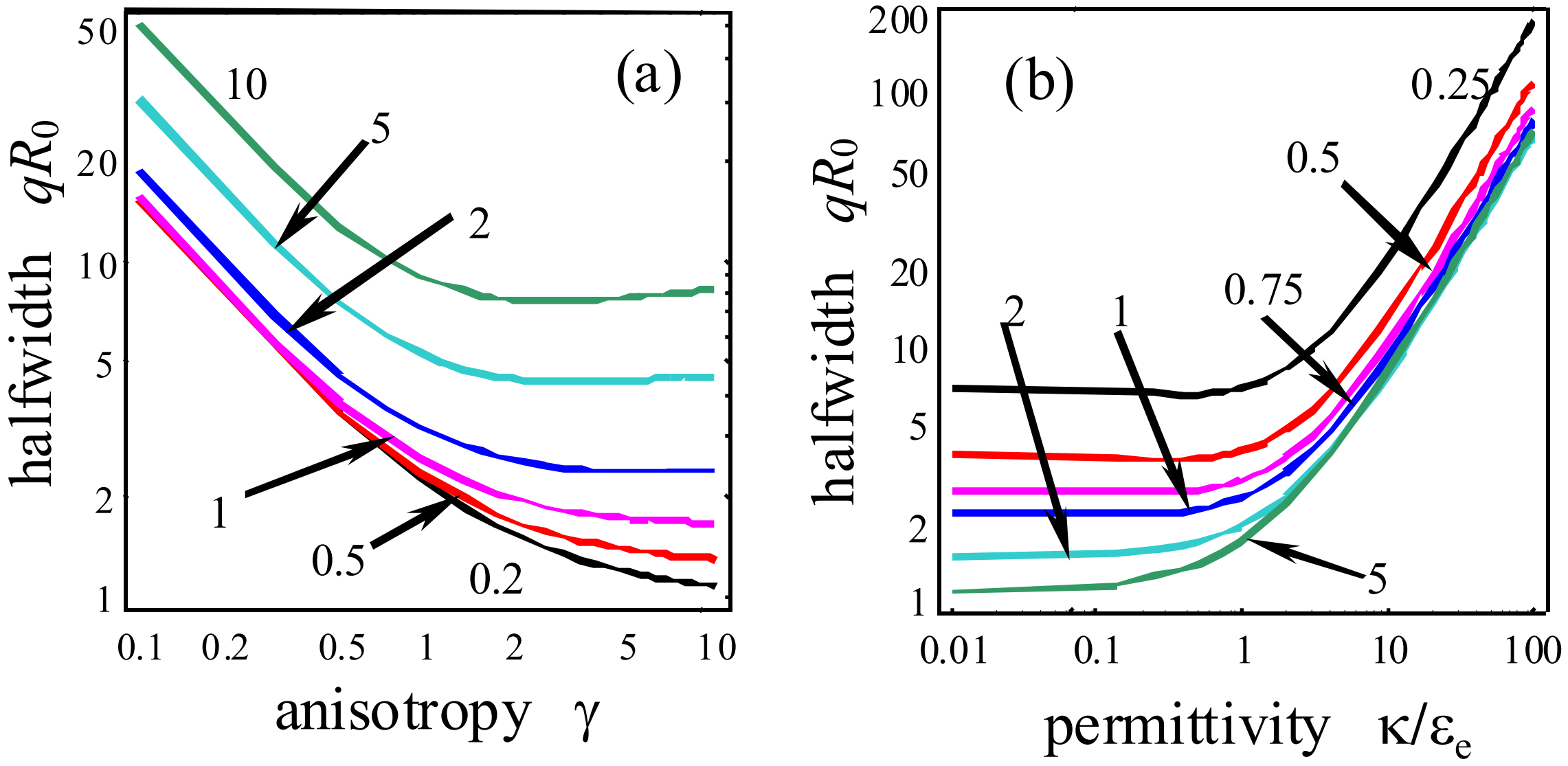

FIG. 6. (Color online). The resolution function component $\widetilde{W}_{333}$ halfwidth vs. the dielectric anisotropy $\gamma$ (a) and relative permittivity $\kappa/\varepsilon_e$ (b) in *q*-space. Figures near the curves correspond to $\gamma$ values (a) and ratio $\kappa/\varepsilon_e$ (b).

### III.5. Effect of material properties on resolution

Note that the lateral resolution $r_{\min}$ is determined by $\rho_{FWHM} \sim 1/q_{FWHM}$ independently on the tip model. Keeping in mind that $d = \varepsilon_e R_0/\kappa$ within the framework of modified point charge model, we obtained the functional dependences:

$$r_{\min} \cong \begin{cases} \dfrac{\gamma\,\varepsilon_e R_0}{2\kappa}\dfrac{1+2\gamma}{(1+\gamma)^2}, & \text{point charge model} \\ \\ \dfrac{\gamma\,\varepsilon_e R_0}{\varepsilon_e+\kappa}\dfrac{1+2\gamma}{(1+\gamma)^2}, & \text{sphere-plane model} \end{cases} \qquad (15)$$

Thus, the functional dependence $r_{\min} \sim \dfrac{\varepsilon_e R_0}{\varepsilon_{11}}$ is valid at $\varepsilon_e \ll \kappa$ for both point charge and sphere-plane models of the tip ($\kappa = \sqrt{\varepsilon_{33}\varepsilon_{11}}$, $\gamma = \sqrt{\varepsilon_{33}/\varepsilon_{11}}$). Hence, it is desirable to decrease external permittivity $\varepsilon_e$ (e.g., in air) and decrease tip radius $R_0$ (atomic tips) in order to increase lateral resolution of PFM, while maintaining good contact. Note, that higher lateral resolution is possible in materials with high $\varepsilon_{11}$ values. The result perfectly explains resolution function halfwidth behavior and curves order in Figs. 3-6, because among the chosen ferroelectric materials $BaTiO_3$ has the highest (1200) and $LiNbO_3$ has the lowest (30) $\varepsilon_{11}$ values.

## IV. Domain Imaging and Domain Wall Profiles

### IV. 1. The response near the flat domain wall

The resolution function and OTF developed in Section III allow calculation of the response from the "point" domain. At the same time, a natural experimental observable in PFM experiment is a domain wall between antiparallel domains. Here, we apply the resolution function theory to the calculation of domain wall profile, derive the approximate expression for domain wall profile in vertical and lateral PFM, and determine the relationship between domain wall width and tip and materials parameters.

Here we calculate surface displacement below the tip located at distance $a$ from the infinitely thin plain domain wall $y_1 = a_0$ ($a_0$ is the domain wall center position) in the point charge approximation of the tip (Fig. 7). The displacement components are given by

$$\begin{aligned} u_i(\mathbf{0},a) = &\int_{-\infty}^{a-a_0} d\xi_1 \int_{-\infty}^{\infty} d\xi_2 \int_0^{\infty} d\xi_3 \frac{\partial G_{ij}(-\xi_1,-\xi_2,\xi_3)}{\partial \xi_k} E_l(\xi_1,\xi_2,\xi_3) c_{kjmn} d_{lnm} - \\ &- \int_{a-a_0}^{\infty} d\xi_1 \int_{-\infty}^{\infty} d\xi_2 \int_0^{\infty} d\xi_3 \frac{\partial G_{ij}(-\xi_1,-\xi_2,\xi_3)}{\partial \xi_k} E_l(\xi_1,\xi_2,\xi_3) c_{kjmn} d_{lnm} \end{aligned} \quad (16)$$

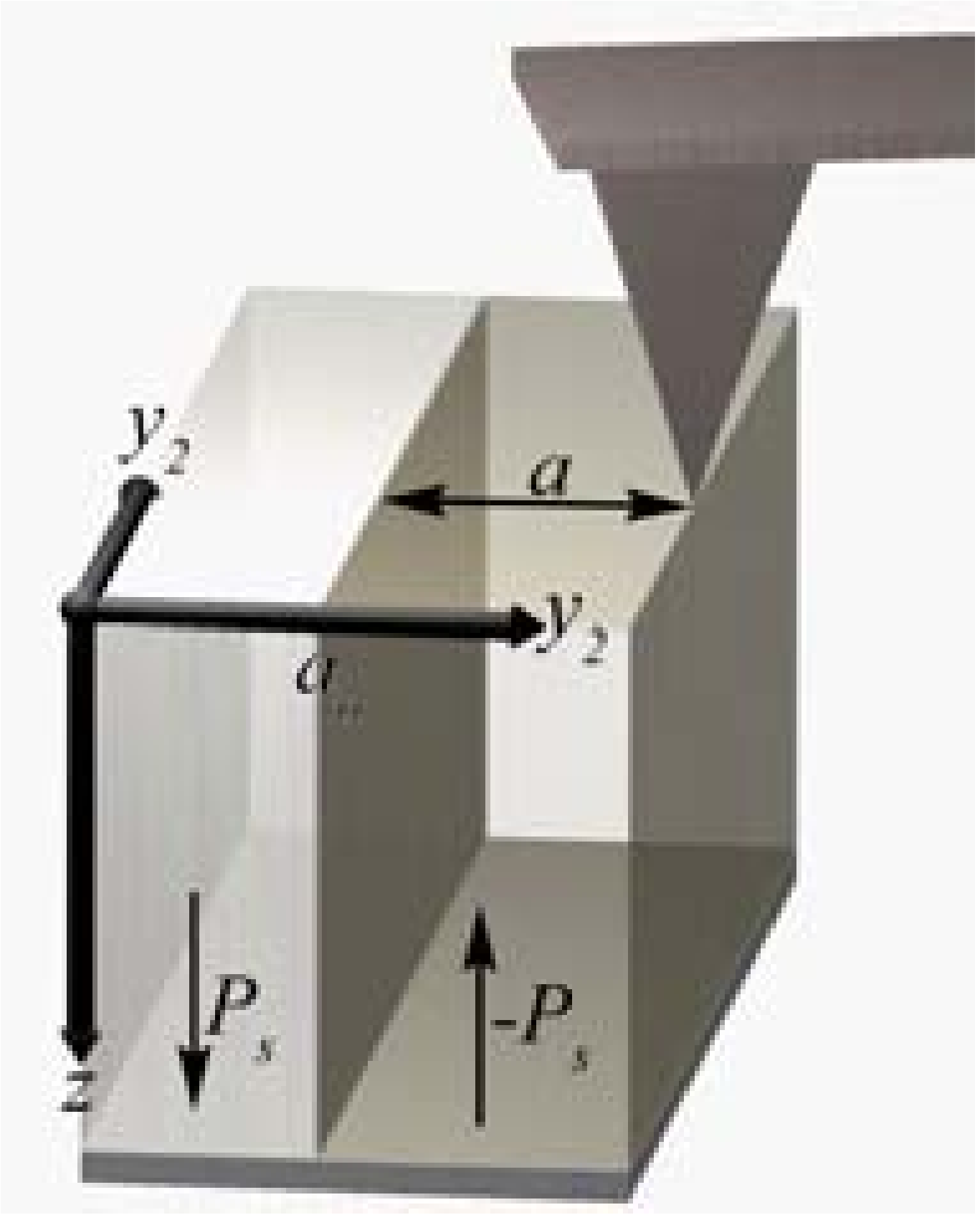

FIG. 7. (Color online). Schematics of PFM measurement across 180° domain wall.

After substitution of corresponding elastic Green's function and electric field distribution, lengthy integration, and Pade analysis, the surface displacement components can be written as:

$$u_1(a) = \frac{1}{2\pi\varepsilon_0(\varepsilon_e+\kappa)}\sum_{m=0}^{\infty}\frac{Q_m}{d_m}\left(g_{113}\left(\frac{|a-a_0|}{d_m},\gamma,\nu\right)d_{31} + g_{151}\left(\frac{|a-a_0|}{d_m},\gamma\right)d_{15} + g_{133}\left(\frac{|a-a_0|}{d_m},\gamma\right)d_{33}\right),$$

$$u_2 = 0$$

$$u_3(a) = \frac{sign(a-a_0)}{2\pi\varepsilon_0(\varepsilon_e+\kappa)}\sum_{m=0}^{\infty}\frac{Q_m}{d_m}\left(g_{313}\left(\frac{|a-a_0|}{d_m},\gamma,\nu\right)d_{31} + g_{351}\left(\frac{|a-a_0|}{d_m},\gamma\right)d_{15} + g_{333}\left(\frac{|a-a_0|}{d_m},\gamma\right)d_{33}\right)$$

(17)

In the case of rigorous sphere-plane model of the tip of curvature $R_0$ that touches the sample $d_m = \frac{R_0}{m+1}$ and $Q_m = \frac{4\pi\varepsilon_0\varepsilon_e U R_0}{m+1}\left(\frac{\kappa-\varepsilon_e}{\kappa+\varepsilon_e}\right)^m$ (e.g., Ref. 30). For modified point charge $Q = 2\pi\varepsilon_0\varepsilon_e R_0 U(\kappa+\varepsilon_e)/\kappa$ located at the distance $d = \varepsilon_e R_0/\kappa$. Exact expressions for $g_{ijk}(s,\gamma,\nu)$ are given in Appendix B.

For both the point charge and sphere-plane models, it is possible to rewrite displacement components in a simple analytical form:

$$u_i = \begin{cases} \dfrac{Q}{2\pi\varepsilon_0(\varepsilon_e+\kappa)d}\, g_{ijk}^{Pade}\left(\dfrac{a-a_0}{d},\gamma,\nu\right)d_{kj}, & \text{point charge model} \\ U\, g_{ijk}^{Pade}\left(\dfrac{a-a_0}{f R_0},\gamma,\nu\right)d_{kj}, & \text{sphere-plane model} \end{cases} \quad (18)$$

where $f=\frac{2\varepsilon_e}{\kappa-\varepsilon_e}\ln\left(\frac{\varepsilon_e+\kappa}{2\varepsilon_e}\right)$. Pade approximations for expressions $g_{ijk}^{Pade}(s,\gamma,\nu)$ are derived in Appendix B. In particular, contributions to vertical PFM signal are

$$g_{351}^{Pade}(s,\gamma)=-\frac{\gamma^2}{(1+\gamma)^2}\cdot\frac{s}{|s|+C_{351}(\gamma)}, \tag{19a}$$

$$g_{333}^{Pade}(s,\gamma)=-\frac{1+2\gamma}{(1+\gamma)^2}\cdot\frac{s}{|s|+C_{333}(\gamma)}, \tag{19b}$$

$$g_{313}^{Pade}(s,\gamma,\nu)=\left(\frac{1+2\gamma}{(1+\gamma)^2}\cdot\frac{s}{|s|+C_{333}(\gamma)}-2\frac{1+\nu}{1+\gamma}\cdot\frac{s}{|s|+C_{313}(\gamma)}\right). \tag{19c}$$

The lateral PFM signal is related to

$$g_{133}^{Pade}(s,\gamma)=\frac{C_{133}(\gamma)}{1+\frac{(1+\gamma)^3}{\gamma}C_{133}(\gamma)|s|}, \tag{19d}$$

$$g_{151}^{Pade}(s,\gamma)=\frac{1}{\frac{1}{2/\pi-C_{133}(\gamma)}+\frac{2(1+\gamma)^3}{(3+\gamma)\gamma^2}|s|}, \tag{19e}$$

$$g_{113}^{Pade}(s,\gamma,\nu)=-\frac{C_{133}(\gamma)}{1+\frac{(1+\gamma)^3}{\gamma}C_{133}(\gamma)|s|}+\frac{(1+\nu)C_{113}(\gamma)}{1+\frac{(1+\gamma)^2}{\gamma}C_{113}(\gamma)|s|}. \tag{19f}$$

The constants $C_{ijk}(\gamma)$ are analytical functions of dielectric anisotropy factor. Here

$C_{351}(\gamma)=\frac{3(1+\gamma)^2}{16\gamma^2}\,{}_2F_1\left(\frac{3}{2},\frac{3}{2};4;1-\frac{1}{\gamma^2}\right)$, in particular $C_{351}(1)=\frac{3}{4}$;

$C_{333}(\gamma)=\frac{3(1+\gamma)^2}{16\gamma^2(1+2\gamma)}\,{}_2F_1\left(\frac{3}{2},\frac{5}{2};4;1-\frac{1}{\gamma^2}\right)$, in particular $C_{333}(1)=\frac{1}{4}$;

$C_{313}(\gamma)=\frac{1+\gamma}{8\gamma^2}\,{}_2F_1\left(\frac{3}{2},\frac{3}{2};3;1-\frac{1}{\gamma^2}\right)$, in particular $C_{313}(1)=\frac{1}{4}$;

$C_{133}(\gamma)=\frac{3}{8\gamma}\cdot {}_2F_1\left(\frac{1}{2},\frac{3}{2};3;1-\frac{1}{\gamma^2}\right)$, in particular $C_{133}(1)=\frac{3}{8}$; $C_{113}(\gamma)=\frac{1}{\gamma}\cdot {}_2F_1\left(\frac{1}{2},\frac{1}{2};2;1-\frac{1}{\gamma^2}\right)$,

in particular $C_{113}(1)=1$. Here ${}_2F_1(p,q;r;s)$ is the hypergeometric function.

Almost exact Pade-exponential tailoring are given in Appendix B, whereas simple Pade approximations (19) are rather good at $s>0.1$, but their rigorously derivatives at $s\to 0$ (i.e., in the center of the domain wall) differ from exact.

When the tip is in contact, piezoresponse signal components $d_{33}^{eff}=u_3(a-a_0)/U$ and $d_{35}^{eff}=u_1(a-a_0)/U$. In the case of weak dielectric anisotropy $\gamma=1$ and $s>0.1$, the domain wall piezo-response signal for the point charge tip model can be found as:

$$d_{33}^{eff}\approx d_{03}+\left[\frac{3}{4}\left(d_{33}+\left(\frac{1}{3}+\frac{4}{3}\nu\right)d_{31}\right)\frac{s}{|s|+1/4}+\frac{1}{4}d_{15}\frac{s}{|s|+3/4}\right] \tag{20a}$$

$$d_{35}^{eff}\approx d_{01}+d_{33}\frac{3/8}{1+3|s|}+d_{31}\left(-\frac{3/8}{1+3|s|}+\frac{1+\nu}{1+4|s|}\right)+d_{15}\frac{2/\pi-3/8}{1+(8/\pi-3/2)|s|}\ . \tag{20b}$$

where $d_{0i}$ is the constant offset related to the electrostatic forces. Again, more rigorous exponential tailoring Pade approximations derived in Appendix B allow better representation of the signal in the center of the wall.

Calculated domain wall profiles for different ferroelectric materials are shown in Figs. 8-9.

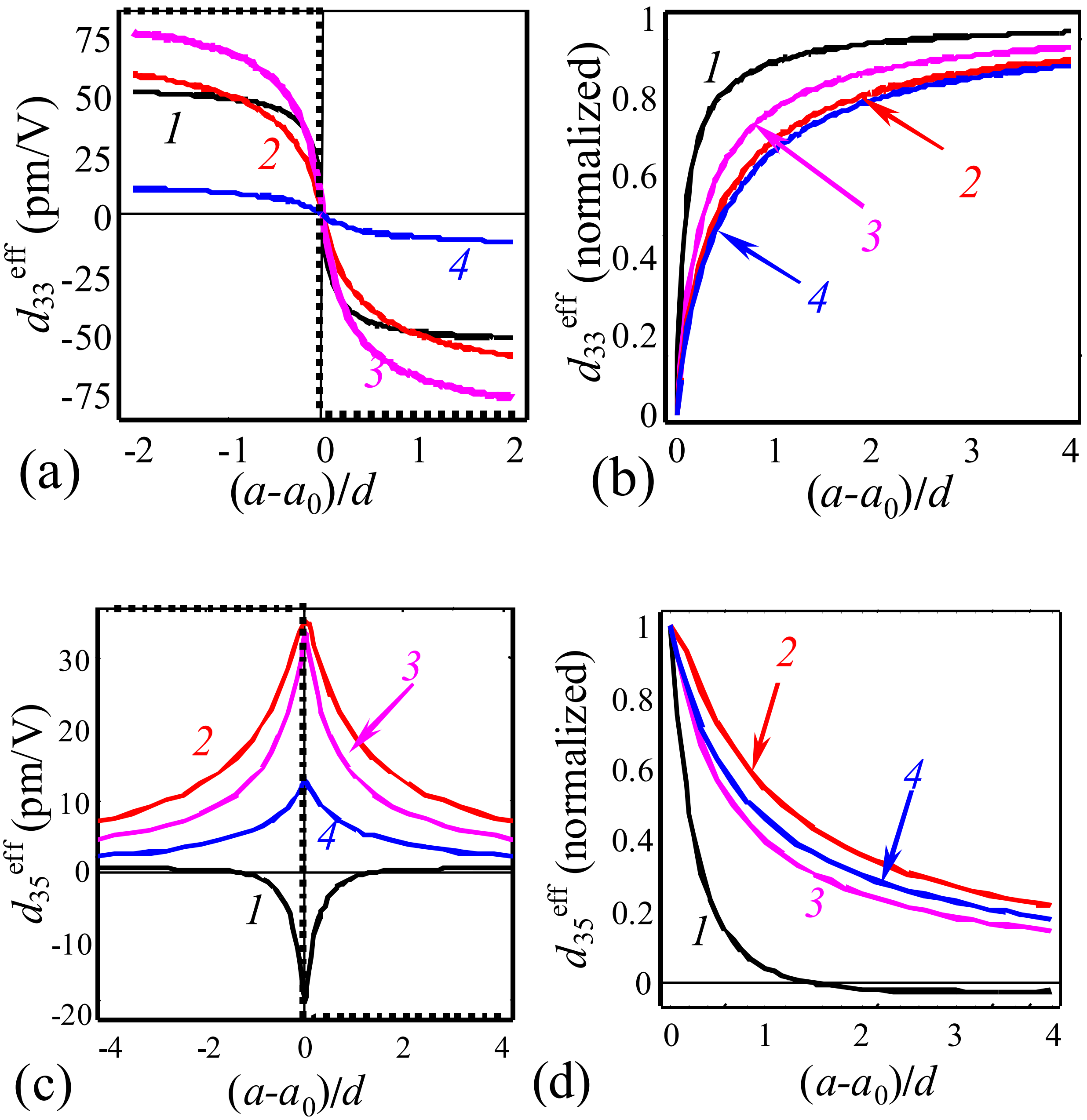

FIG. 8. (Color online). Domain wall piezo-response profile for point charge model of the tip at $\nu = 0.35$ for different ferroelectric materials $BaTiO_3$ (1), PZT6B (2), $PbTiO_3$ (3), $LiNbO_3$ (4). Shown is (a) vertical PFM signal, (b) normalized vertical PFM signal, (c) lateral PFM signal, and (d) normalize lateral PFM signal.

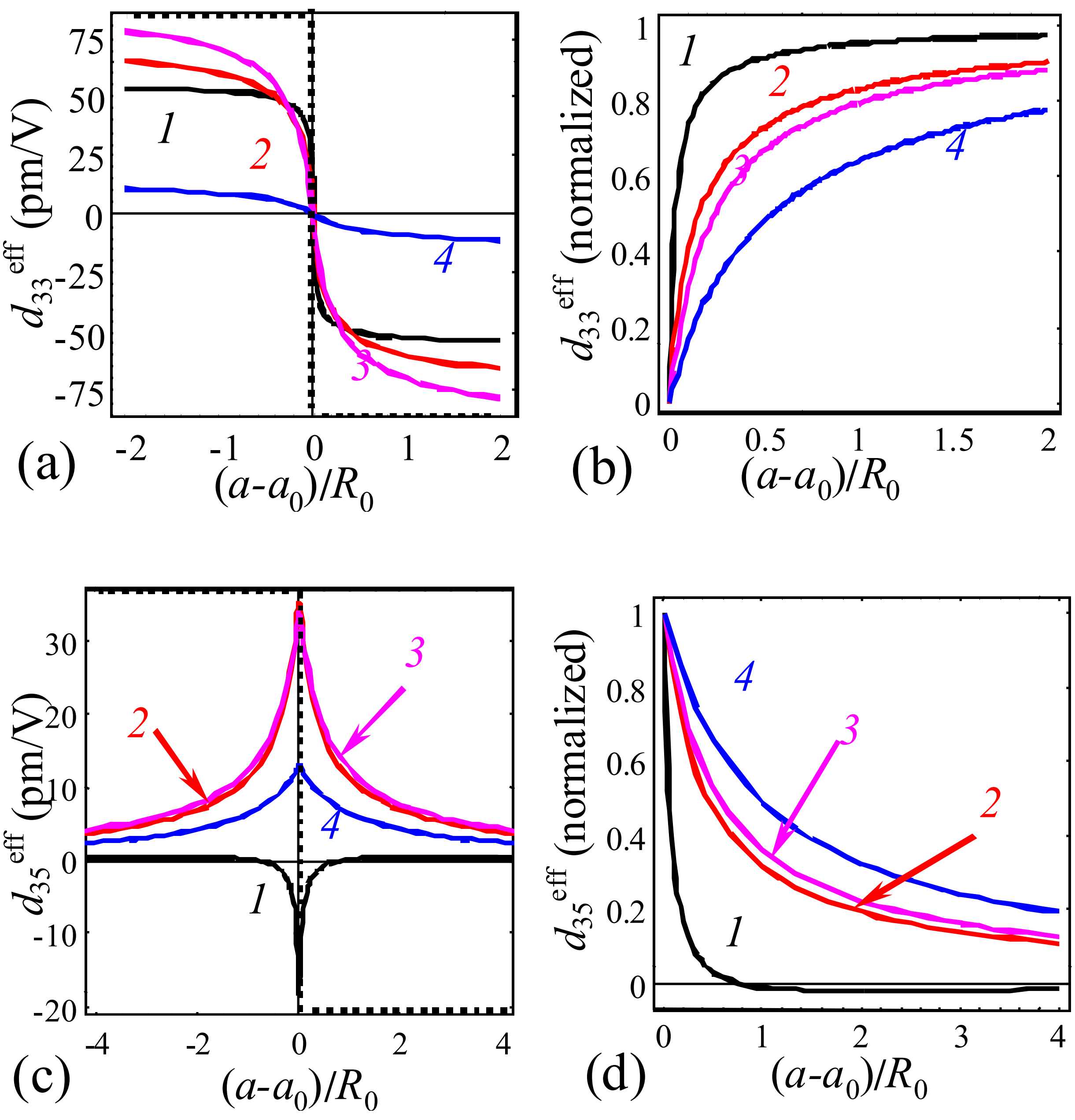

FIG. 9. (Color online). Domain wall piezo-response profile for sphere-plane model of the tip at $\nu = 0.35$ for different ferroelectric materials $BaTiO_3$ (1), PZT6B (2), $PbTiO_3$ (3), $LiNbO_3$ (4). Shown is (a) vertical PFM signal, (b) normalized vertical PFM signal, (c) lateral PFM signal, and (d) normalize lateral PFM signal.

Note, that the sphere-plane tip gives more "rectangular" image of the ideal domain wall in comparison with the sloped one given by the point charge tip for the same values of dimensionless distance $s$ (i.e., for $d = R_0$) [compare Fig. 9 (a,c) and 9 (a,c)]. Therefore, sphere-plane tip has higher lateral resolution in comparison with the point charge one for comparable characteristic dimensions. This behavior is anticipated due to the concentration of charges at the tip-surface junction in sphere-plane model.

It also follows from Figs. 8-9, that the best lateral resolution can be achieved in $BaTiO_3$, whereas the lowest corresponds to the $LiNbO_3$ independent on the tip representation [compare curve order in plots (b) and (d)]. This is in agreement with Eq. (10), since from the chosen ferroelectric materials $BaTiO_3$ has the highest and $LiNbO_3$ has the lowest $\varepsilon_{11}$ values. The situation for $d_{35}^{eff}$ is more complex: here the resolution for $BaTiO_3$ is the highest one, but the signal change sign, since the negative contribution of $d_{15}$ dominates far from the domain wall. The lowest resolution corresponds to PZT6B.

### IV. 2. Resolution effect on imaging single domain structures

The resolution effects on imaging single- and double domain structures is illustrated in Fig. 10 (a,b) and (c,d) correspondingly for the point charge representation of the tip.

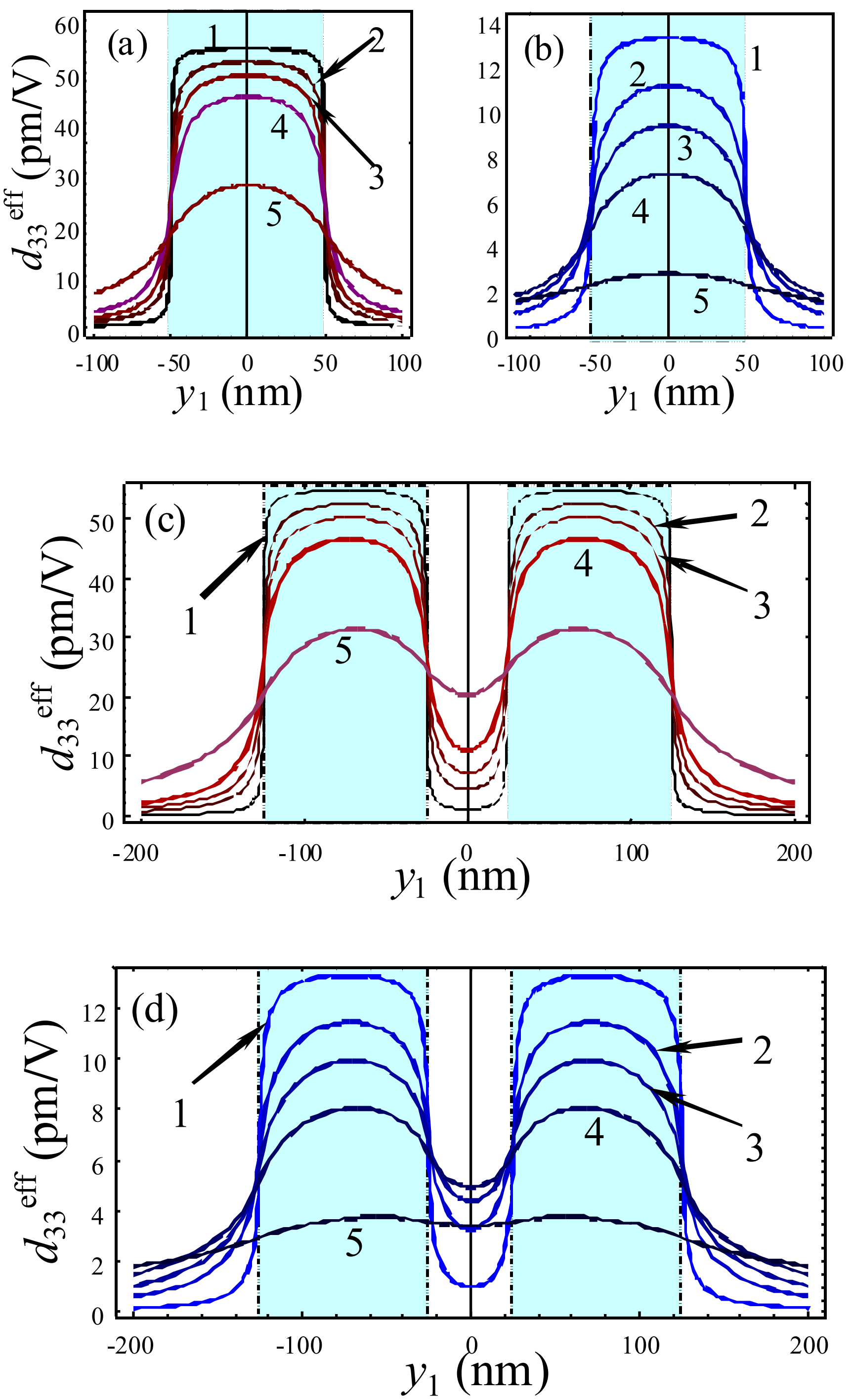

FIG. 10. (Color online). Effective piezoresponse $d_{33}^{eff}$ of one domain ((a) - $BaTiO_3$, (b) - $LiNbO_3$) and two domains [(c) - $BaTiO_3$, (d) - $LiNbO_3$)] vs. the distance $y_1$ for different charge-surface separation $d = 5\,nm$ (curves 1), $d = 25\,nm$ (curves 2), $d = 50\,nm$ (curves 3),

$d = 100nm$ (curves 4), $d = 500nm$ (curves 5) and fixed ratio $Q/d$ (i.e., constant potential $U$). Filled regions designate the domains.

The effect of the charge-surface separation distance $d$ on domain contrast is demonstrated for ferroelectrics $BaTiO_3$ and $LiNbO_3$. It is clear that for small $d$ values the piezoresponse amplitude is high and the signal shape approaches real domain shape. The signal diffuses and resolution strongly decreases under the distance $d$ increase. The effect is the more pronounced for the two-point resolution in $LiNbO_3$, when the domains become unresolved at high $d$ values (curves 4,5). However, the signal never achieves zero value and the condition of the *observability* of the domain is determined by the noise level of the system (thermal noise) and cross-talk with topographic inhomogeneities (frozen disorder).

The maximal contrast $d_{33}^{\max}$ in the center of a domain vs. charge-surface separation $d/a$ for single domain stripe and periodic domain structure is shown in Fig. 11 (a). Note, that the ratio $Q/d$ is fixed as proportional to the tip potential $U$, thus the law $d_{33}^{\max} \sim 1/d$ is valid at high $d/a >> 1$ [see Eq. (11)]. The resolution decreases under the separation increase.

Note that with the decrease of domain size (comp. Fig. 11 b, c), the signal in the center of the domain decreases. The observability of the single domain is determined by the condition that the signal at the center is above the noise level of the system, as described above. Similar behavior is anticipated from the domain of opposite polarity, as illustrated in Fig. 12 b,c. In all cases, the highest resolution is anticipated for $BaTiO_3$, whereas the lowest one corresponds to the $LiNbO_3$.

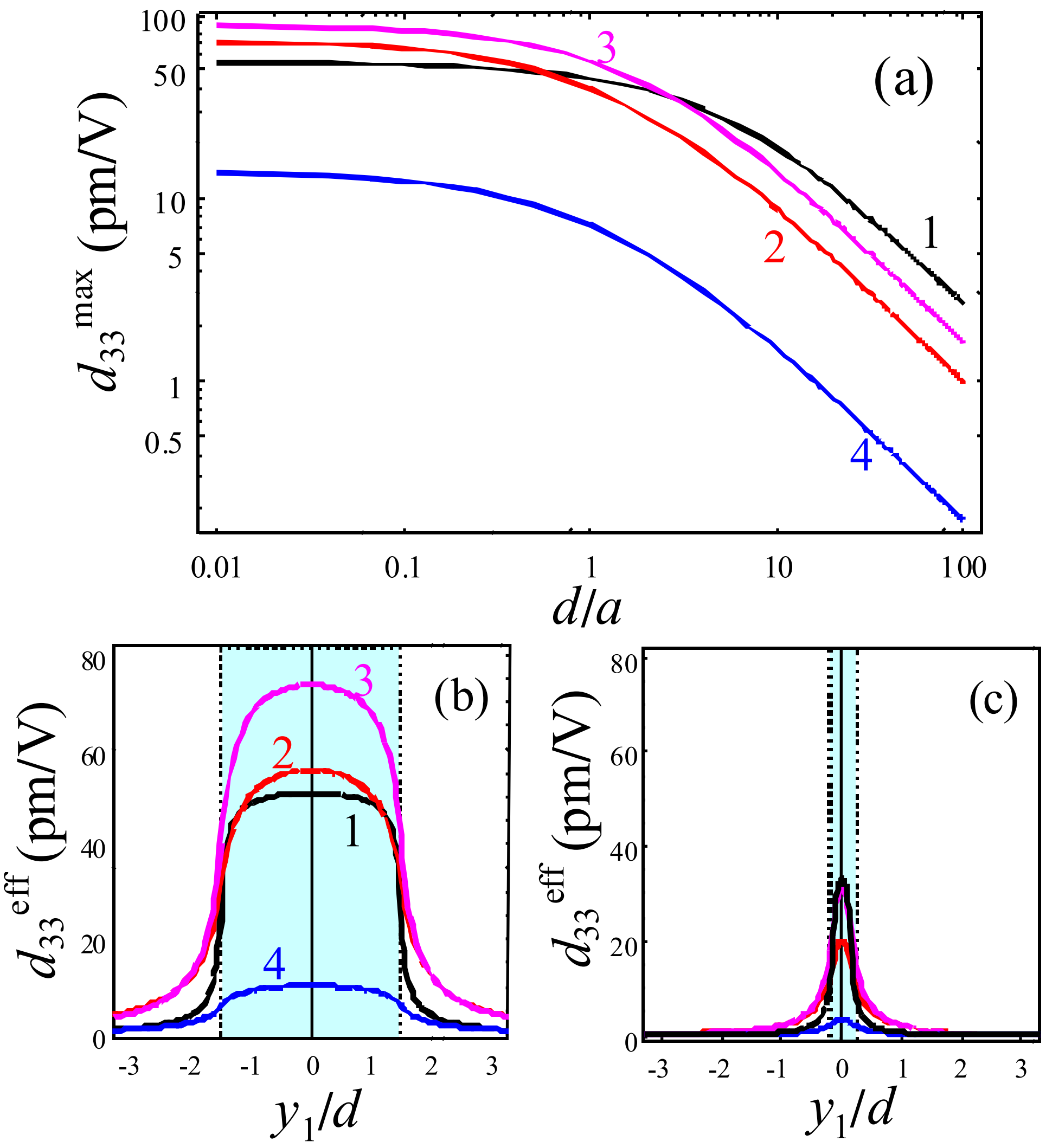

FIG. 11. (Color online). The maximal contrast $d_{33}^{max}$ in the center of a domain vs. charge-surface separation $d/a$ for single domain stripe (a). Effective piezoresponse $d_{33}^{eff}$ of one domain (b,c) vs. the distance $y_1$ for different domain width. The fixed ratio $Q/d$ is proportional to the tip potential $U$. Curves correspond to the different ferroelectric materials $BaTiO_3$ (curves 1), PZT6B (curves 2), $PbTiO_3$ (curves 3), $LiNbO_3$ (curves 4). Filled region designate the domain width (ideal image).

### IV. 3. Resolution effect on imaging periodic domain structures

Similar analysis can be derived in the context of periodic domain structures (e.g., lamellar domains or nanodomains in ultrathin ferroelectric films).[46] Here, we derive the signal strength using analytical form for the transfer function derived in Section III. Let us consider lateral resolution of the perfect periodic $180^{o}$ domain structure (ideal rectangular wave). The variation of piezoelectric properties (ideal image) can be represented as a Fourier series:

$$d_{klj}(\mathbf{y}) = d_{klj}\sum_{n=0}^{\infty}\frac{4}{(2n+1)\pi}\sin\left(\frac{2\pi}{a}(2n+1)y_1\right), \tag{21}$$

where $a$ is the domain width. The experimental image in the Fourier domain is then

$$u_3(a,y_1) = \sum_{n=0}^{\infty}\frac{4}{(2n+1)\pi}\sin\left(\frac{2\pi}{a}(2n+1)y_1\right)F_q\left(\frac{2\pi}{a}(2n+1)\right), \tag{22}$$

where the object transfer function is given by integration of Eq. (11a) as evaluated at the allowed wave vectors, $q_1 = 2\pi(2n+1)/a$ and $q_2 = 0$.

Note that for low spatial resolutions, only the first harmonic component is transferred to the image, and experimental image will have the functional form

$$u_3(y) = \frac{4}{\pi}\sin\left(\frac{2\pi y_1}{a}\right)F_q\left(\frac{2\pi}{a}\right), \tag{23}$$

i.e., sinusoidal signal with amplitude decreasing with domain size [comp. Fig. 2 (c)].

The maximal contrast $d_{33}^{\max}$ in the center of a domain vs. charge-surface separation $d/a$ for periodic domain structure is shown in Fig. 12 (a). Similarly to the case of single domain stripe (see Fig. 11a) the law $d_{33}^{\max} \sim 1/d$ is valid only at $d/a >> 1$ [see Eq. (11)]. Insets in plot (b) show effective piezoresponse $d_{33}^{eff}$ of periodic domain structure vs. the distance $y_1$ for different charge-surface separation $d$. The resolution decreases under the separation increase.

It is clear that for low resolution (e.g., $d/a \geq 1$) only first harmonic $\sim \sin(2\pi y_1/a)$ survive instead of the rectangular wave.

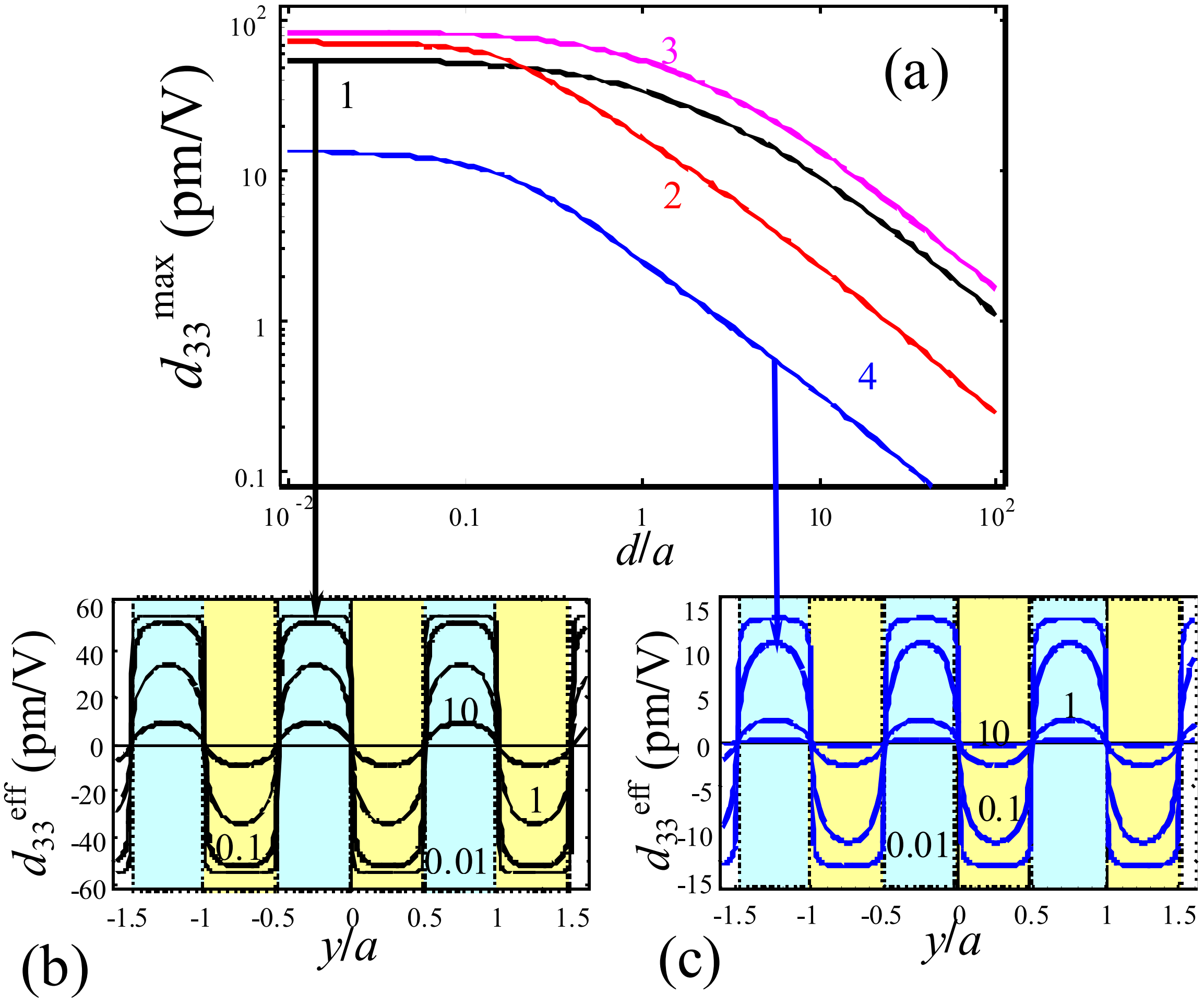

FIG. 12. (Color online). The maximal contrast $d_{33}^{max}$ in the center of a domain vs. charge-surface separation $d/a$ for a periodic domain structure (a). Curves correspond to the different ferroelectric materials $BaTiO_3$ (curves 1), PZT6B (curves 2), $PbTiO_3$ (curves 3), $LiNbO_3$ (curves 4). Insets (b,c) show effective piezoresponse $d_{33}^{eff}$ of periodic domain structure vs. the distance $y_1$ for different charge-surface separation $d/a$=0.01; 0.1; 1; 10 (figures near the curves) in $BaTiO_3$ (b), and $LiNbO_3$ (c). Filled regions designate the domains.

## V. The Response of Cylindrical Domains

One of the key problems in the interpretation of PFM spectroscopy is establishing the relationship between dc-bias dependent response and the size of formed domain below the tip. In an elegant work, Kholkin et al.,[47] has demonstrated concurrent hysteresis loop measurements and acquisition of PFM image. However, such experiments, while providing unambiguous information on the size of domain on the different stages of switching process, are extremely time consuming and tedious. Moreover, the imaging time ($\sim 10^2 - 10^3$ s) is significantly larger than the time resolution of single-point PFM measurement ($10^{-3}$ s), allowing for multiple relaxation processes after switching. Quantitative spectroscopic PFM measurements suggest a strategy to avoid these limitations. However, data interpretation requires quantitative relationship between the geometric parameters of the grown domain below the tip and PFM signal to be established.

Domain formed in semi-infinite material below the tip is characterized by at least two geometric parameters, domain radius and length, while measured PFM signal is a single variable. Hence, in the most general case, domain geometric parameters of the domain could not be extracted unambiguously from the PFM data. However, the domain is typically elongated (i.e., highly anisotropic needle-type domains are formed both in classical and PFM geometry).[48] At the same time, the electrostatic field generated by the PFM tip is concentrated in the near-surface layer, as dictated by the dielectric anisotropy of material.[30,31] Hence, *domain radius* is a primary parameter defining PFM signal, while domain length is can be approximated as infinite except for the early stages of domain growth process.

## V. 1. The response of the single cylindrical domain

Hence, interpretation of PFM spectroscopy data can be achieved using a model of a semi-infinite cylindrical domain. Here we assume that domain wall is purely cylindrical and the charged tip is located in the domain center (0,0,0) (Fig. 13).

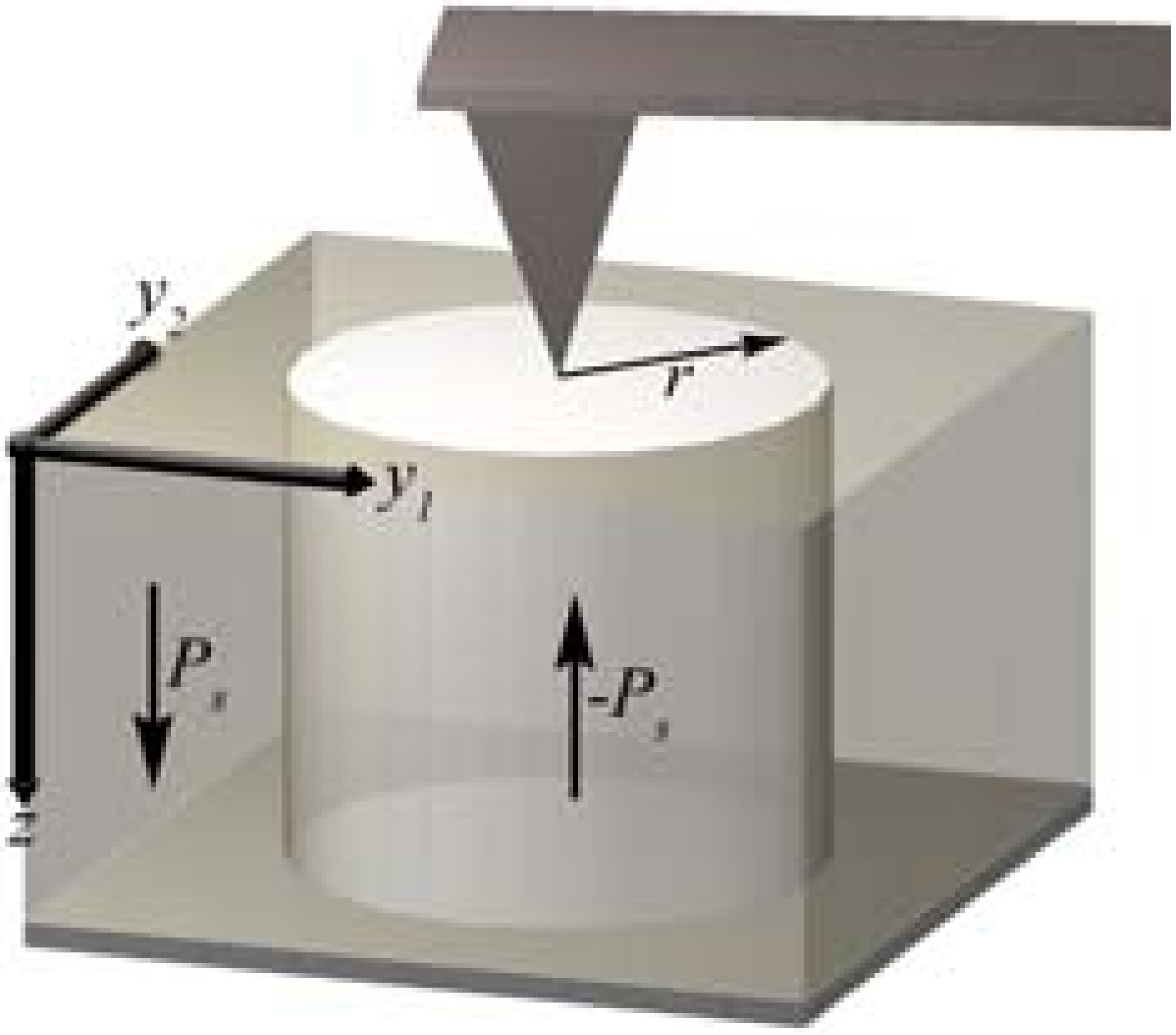

FIG.13 (Color online). Response calculation from cylindrical domain.

The displacement in the center of a domain acquires the form:

$$u_i(\mathbf{0},r)=2\pi\left(\int_0^{\infty}\rho d\rho\int_0^{\infty}dz\frac{\partial G_{ij}(\boldsymbol{\rho},z)}{\partial\xi_k}E_l(\rho,z)-2\int_0^{r}\rho d\rho\int_0^{\infty}d\xi_3\frac{\partial G_{ij}(\boldsymbol{\rho},z)}{\partial\xi_k}E_l(\rho,z)\right)c_{kjmn}d_{lnm}, \quad (25)$$

where $\sqrt{\xi_1^2+\xi_2^2}=\rho$ and $\xi_3=z$ (see Appendix C for details). Using cylindrical symmetry of the problem, substituting the correct form for elastic Green's function and electrostatic field, integration and Pade analysis leads to the following equations for surface displacement:

$$u_3(r) = \frac{1}{2\pi\varepsilon_0(\varepsilon_e + \kappa)} \sum_{m=0}^{\infty} \frac{Q_m}{d_m} \left( h_{13}\left(\frac{r}{d_m}, \gamma, \nu\right) d_{31} + h_{51}\left(\frac{r}{d_m}, \gamma\right) d_{15} + h_{33}\left(\frac{r}{d_m}, \gamma\right) d_{33} \right), \qquad (26)$$

while $u_1 = u_2 = 0$ as follows from the symmetry considerations. Exact expressions for the functions $h_{ij}(s,\gamma,\nu)$ are given in Appendix C.

Both for the point charge and sphere-plane models the displacements can be written in a simple analytical form:

$$u_3 = \begin{cases} \dfrac{Q}{2\pi\varepsilon_0(\varepsilon_e + \kappa)d} h_{jk}^{Pade}\left(\dfrac{r}{d}, \gamma, \nu\right) d_{kj}, & \text{point charge model} \\[2ex] U h_{jk}^{Pade}\left(\dfrac{r}{f R_0}, \gamma, \nu\right) d_{kj}, & \text{sphere} - \text{plane model} \end{cases} \qquad (27)$$

Pade approximations $h_{jk}^{Pade}(s,\gamma,\nu)$ for expressions $h_{jk}(s,\gamma,\nu)$ are derived in Appendix C, namely:

$$h_{33}^{Pade}(s,\gamma) \approx -\frac{1+2\gamma}{(1+\gamma)^2} + 2\frac{1+2\gamma}{(1+\gamma)^2} \cdot \frac{s}{s + D_{33}(\gamma)}, \qquad (28a)$$

$$h_{13}^{Pade}(s,\gamma,\nu) \approx \frac{1+2\gamma}{(1+\gamma)^2} - \frac{2(1+\nu)}{1+\gamma} - 2\left( \frac{1+2\gamma}{(1+\gamma)^2} \cdot \frac{s}{s + D_{33}(\gamma)} - \frac{2(1+\nu)}{1+\gamma} \cdot \frac{s}{s + D_{13}(\gamma)} \right). \qquad (28b)$$

$$h_{51}^{Pade}(s,\gamma) \approx -\frac{\gamma^2}{(1+\gamma)^2} + 2\frac{\gamma^2}{(1+\gamma)^2} \cdot \frac{s^2}{2\dfrac{\gamma^2}{(1+\gamma)^2} + D_{51}(\gamma)\, s + s^2}, \qquad (28c)$$

Constants $D_{ij}(\gamma)$ depend solely on the dielectric anisotropy of material and

$D_{51}(\gamma) = \dfrac{3\pi(1+\gamma)^2}{32\gamma^4}\left[ -2\gamma^2 {}_2F_1\left(\dfrac{1}{2}, \dfrac{3}{2}; 3; 1 - \dfrac{1}{\gamma^2}\right) + {}_2F_1\left(\dfrac{3}{2}, \dfrac{3}{2}; 4; 1 - \dfrac{1}{\gamma^2}\right)\right]$, in particular $D_{51}(1) = \dfrac{3\pi}{8}$;

$D_{33}(\gamma) = \dfrac{3\pi(1+\gamma)^2}{32\gamma^2(1+2\gamma)} \cdot {}_2F_1\left(\dfrac{3}{2}, \dfrac{5}{2}; 4; 1 - \dfrac{1}{\gamma^2}\right)$, in particular $D_{33}(1) = \dfrac{\pi}{8}$;

$D_{13}(\gamma)=\pi\frac{1+\gamma}{16\gamma^2}\,{}_2F_1\left(\frac{3}{2},\frac{3}{2};3;1-\frac{1}{\gamma^2}\right)$, in particular $D_{13}(1)=\frac{\pi}{8}$. Here ${}_2F_1(p,q;r;s)$ is the hypergeometric function.

Much more precise but cumbersome Pade-exponential approximations for expressions $h_{jk}(s,\gamma,\nu)$ are given in Appendix C. Note, that exact curves almost coincide with Pade-exponential tailoring (17), whereas simple Pade approximations are rather good at $s>0.1$, but have non-rigorous derivatives at $s\to 0$ (see Appendix C for details).

For good tip-surface contact, piezoresponse signal is $d_{33}^{eff}=u_3(r)/U$ and from Eqs. (26, 27) can be written as:

$$d_{33}^{eff}=d_0+h_{13}d_{31}+h_{51}d_{15}+h_{33}d_{33}\ , \tag{29}$$

where $d_0$ is the instrumental offset. For materials with weak dielectric anisotropy $\gamma=1$, PFM signal for point charge tip model can be fitted as:

$$d_{33}^{eff}\approx d_0-\left[\frac{3}{4}\left(d_{33}+\left(\frac{1}{3}+\frac{4}{3}\nu\right)d_{31}\right)\frac{\pi d-8r}{\pi d+8r}+\frac{d_{15}}{4}\frac{3\pi d-8r}{3\pi d+8r}\right]. \tag{30}$$

Relative piezo-response $\Delta d_{33}^{eff}(r)=\left[d_{33}^{eff}(r)-d_{33}^{eff}(0)\right]/2$ vs. the cylindrical domain radius for point charge model and sphere-plane model of the tip from different ferroelectric materials is shown in Figs. 14-15. Note, that $\Delta d_{33}^{eff}(r)$ is defined in such way to coincide with the response from a single piezoelectric cluster embedded into non-piezoelectric matrix, as opposed to the definition of the signal between antiparallel domains.

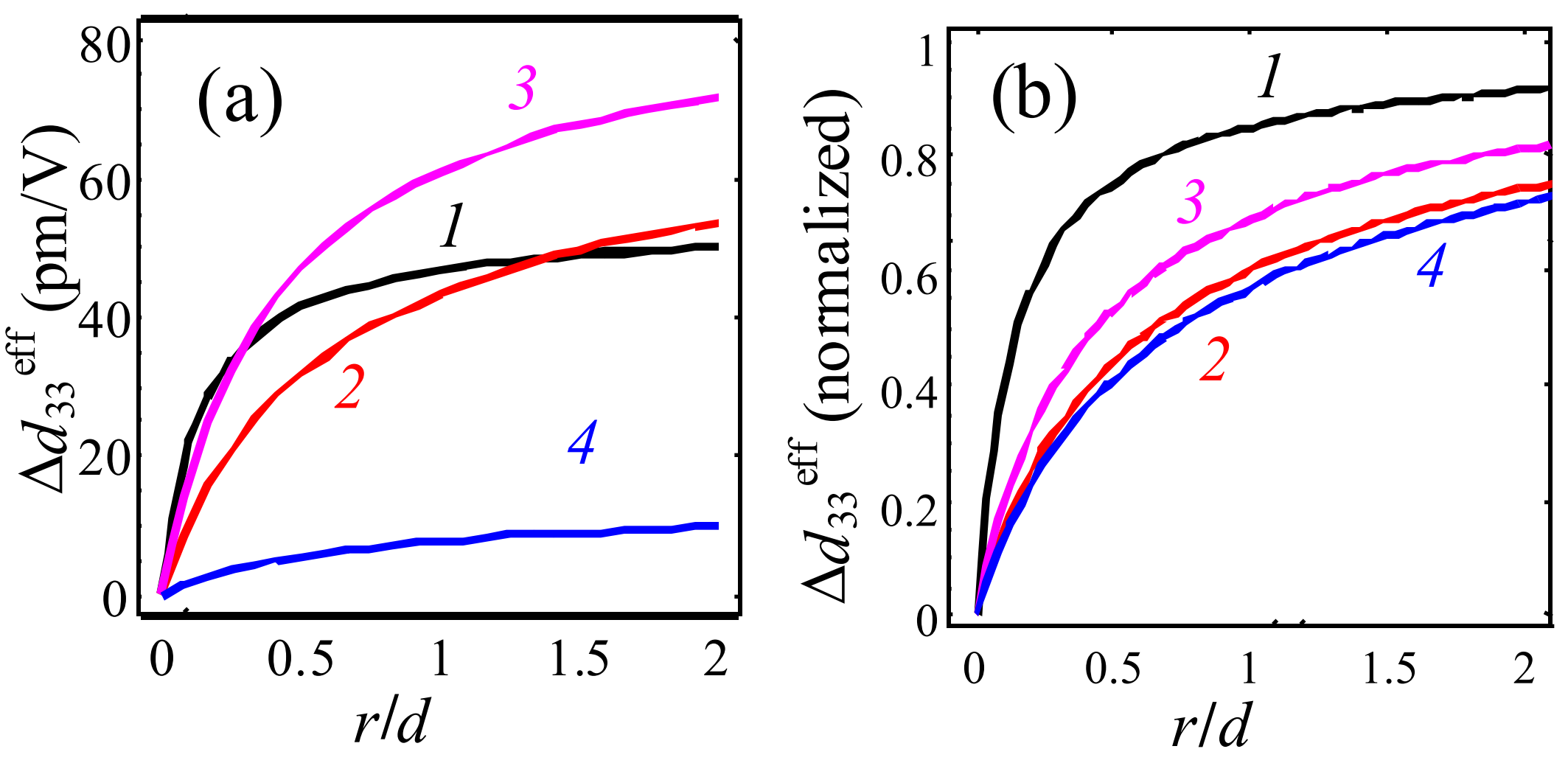

FIG. 14 (Color online). Relative piezo-response $\Delta d_{33}^{eff}(r)$ vs. the cylindrical domain radius for point charge model of the tip at $\nu = 0.35$ for different ferroelectric materials $BaTiO_3$ (1), PZT6B (2), $PbTiO_3$ (3), $LiNbO_3$ (4). Shown are (a) absolute and (b) normalize values.

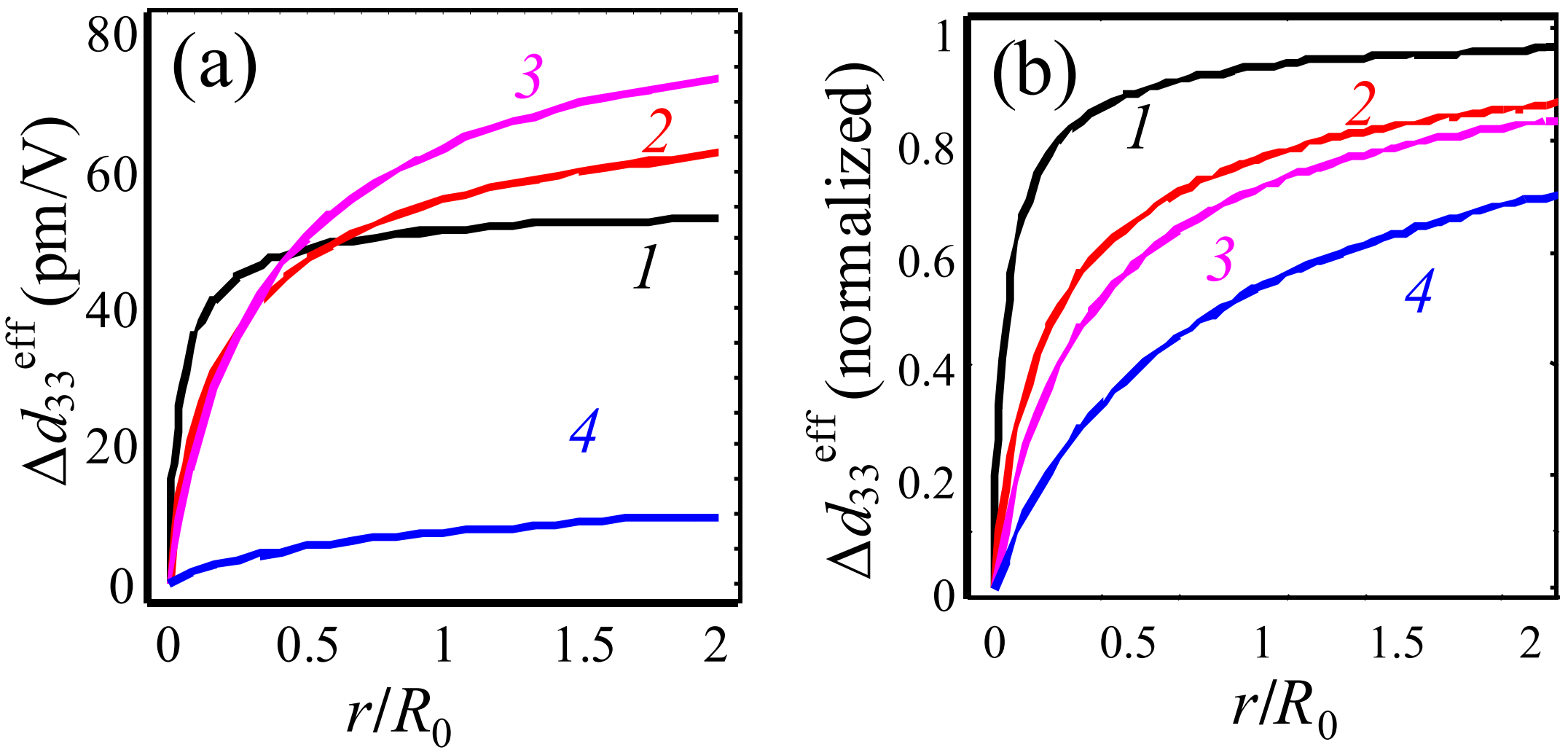

FIG. 15 (Color online). Domain wall piezo-response profile for sphere-plane model of the tip at $\nu = 0.35$ for different ferroelectric materials $BaTiO_3$ (1), PZT6B (2), $PbTiO_3$ (3), $LiNbO_3$ (4). Shown are (a) absolute and (b) normalize values.

Note, that similarly to stripe domain case, the sphere-plane model predicts higher sensitivity to small domains compared to point charge tip at the same values of dimensionless distance, $s$ (i.e., at $d = R_0$) [compare Fig. 15 (a) and (b)]. Also it follows from Figs. 14-15, that the best sensitivity to small domains formed below the tip can be achieved in $BaTiO_3$, whereas the worst one corresponds to the $LiNbO_3$ independently on the tip representation. The results are in agreement with the predictions of Eq. (13), because among the chosen ferroelectric materials $BaTiO_3$ has the highest and $LiNbO_3$ has the lowest $\varepsilon_{11}$ values.

### V. 2. Piezoelectric response of nested domains and ferroelectric tubes

Domain formation induced by bias applied to AFM tip proceeds through several stages of nucleation and growth.[26] As discussed in Section V.1, the domain walls can be approximated as cylindrical near the sample surface. For both thermodynamic and weak pinning limits, the initial domain nucleation occurs at critical voltage $-U_{cr}$.[25] Then domain radius increases under the further voltage increase. Under the voltage decrease, the domain wall is pinned by lattice and defects, precluding shrinking[26] For negative voltages, a sufficiently "big" domain acts as new matrix for a new domain appearing just below the tip at $U \leq -U_{cr}$, i.e. nested domains (tube) form. The tube thickness decrease under the further voltage increase and finally domain walls annihilate and the system returns to the initial state. In thermodynamic limit the radius of cylindrical domain decreases with voltage decrease, no nested domains occur. Thus, in order to calculate domain hysteresis allowing for pinning, one has to derive expressions for domain tube piezoresponse.

Similar problem arises in the context of ferroelectric nanowires and nanotubes.[49] It has been shown that ferroelectric nanotubes posses remnant polarization and exhibit ferroelectric hysteresis.[50] Morrison *et al* [51] demonstrated that ultra-small $PbZr_{0.52}Ti_{0.48}O_3$ nanorods and nanotubes possess rectangular shape of the piezoelectric hysteresis loop with effective remnant piezoelectric coefficient value compatible with the ones typical for PZT films. Also the authors demonstrated that the ferroelectric properties of the free $BaTiO_3$ nanotubes are perfect. Poyato et al.[52] with the help of piezoelectric response force microscopy found that nanotube-patterned BaTiO3 film reveals ferroelectric properties. Also they demonstrated the existence of local piezoelectric and oriented ferroelectric responses, prior to the application of a dc field. Thus, in order to simulate ferroelectric nanotubes polar properties, one has to derive relevant expressions for their piezoelectric response.

Hereinafter we assume that domain walls are purely cylindrical and the charged tip is located in the tube(s) center that is typical for domain reversal (see Fig. 13). Using cylindrical symmetry of the problem, it is easy to extend analyses of section V.2, namely the piezoresponse of inverted nested domains, or ferroelectric nanotubes embedded in non-polar matrix, can be represented as a superposition of responses (26)-(28). Thus, both for the point charge and sphere-plane models these displacements can be written in a simple analytical form.

Displacement in the center of inverted domain tube in ferroelectric matrix:

$$u_3(r_o,r_i)=\begin{cases}\dfrac{Q\cdot d_{kj}}{2\pi\varepsilon_0(\varepsilon_e+\kappa)d}\left(h_{jk}^{Pade}(0,\gamma,\nu)-h_{jk}^{Pade}\left(\dfrac{r_o}{d},\gamma,\nu\right)+h_{jk}^{Pade}\left(\dfrac{r_i}{d},\gamma,\nu\right)\right), & \text{point charge model}\\ U\cdot d_{kj}\left(h_{jk}^{Pade}(0,\gamma,\nu)-h_{jk}^{Pade}\left(\dfrac{r_o}{f\,R_0},\gamma,\nu\right)+h_{jk}^{Pade}\left(\dfrac{r_i}{f\,R_0},\gamma,\nu\right)\right), & \text{sphere}-\text{plane model}\end{cases} \quad (31)$$

While $u_1=u_2=0$ as follows from the symmetry considerations; $r_o$ is the tube outer radius, $r_i$ is the inner one. Expressions for the functions $h_{jk}^{Pade}(s,\gamma,\nu)$ are given by Eqs.(28). Effective piezo-response profile in the centre of domain ring vs. outer radius $r_o/d$ is shown in Fig.16 for different ferroelectric materials.

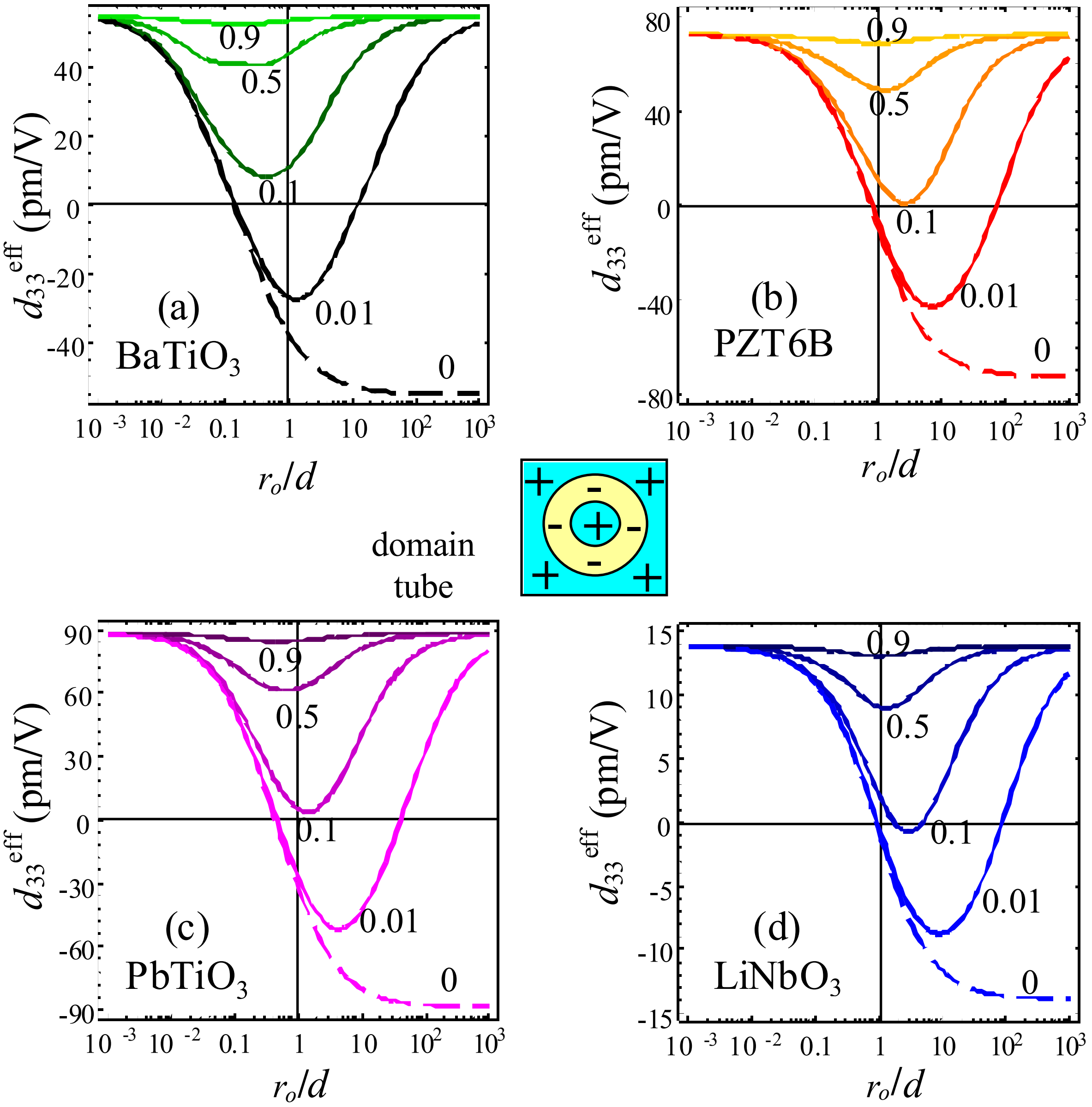

FIG. 16 (Color online). Effective piezo-response profile in the center of domain tube vs. dimensionless outer radius $r_o/d$ for point-charge model of the tip for different ferroelectric materials $BaTiO_3$ (a), PZT6B (b), $PbTiO_3$ (c), $LiNbO_3$ (d). Other parameters: $\nu = 0.35$; figures near the curves corresponds to the ratio $r_i/r_o = 0; 0.01; 0.1; 0.5; 0.9$.

From the data in Fig. 16, the piezo-responses of cylindrical domain and nanotube embedded in non-polar matrix become completely different at $r_o/d>1$ and $r_i/r_o\geq 0.01$ (compare solid and dashed curves). For the tube with $r_i/r_o\geq 0.1$ the response does not change the sign, however the pronounced minimum appeared at $r_o/d\sim 1$. The minimum depth decreases under the ratio $r_i/r_o$ increase and disappears at $r_i/r_o\to 1$ as it should be expected for the uniformly polarized matrix.

Displacement in the center of ferroelectric tube (ring cluster in nonpolar matrix) is:

$$u_3(r_o,r_i)=\begin{cases}\dfrac{Q\cdot d_{kj}}{4\pi\varepsilon_0(\varepsilon_e+\kappa)d}\left(h_{jk}^{Pade}\left(\dfrac{r_o}{d},\gamma,\nu\right)-h_{jk}^{Pade}\left(\dfrac{r_i}{d},\gamma,\nu\right)\right), & \text{point charge model}\\[2ex] \dfrac{U\cdot d_{kj}}{2}\left(h_{jk}^{Pade}\left(\dfrac{r_o}{f\,R_0},\gamma,\nu\right)-h_{jk}^{Pade}\left(\dfrac{r_i}{f\,R_0},\gamma,\nu\right)\right), & \text{sphere}-\text{plane model}\end{cases}\qquad(32)$$

It is clear that the response (32) differs from (31) on constant value $h_{jk}^{Pade}(0,\gamma,\nu)$ and factor of 2. Displacement in the center of filled ferroelectric tube (nested ferroelectric clusters in nonpolar matrix) has the form:

$$u_3(r_o,r_i)=\begin{cases}\dfrac{Q\cdot d_{kj}}{4\pi\varepsilon_0(\varepsilon_e+\kappa)d}\left(h_{jk}^{Pade}\left(\dfrac{r_o}{d},\gamma,\nu\right)-2h_{jk}^{Pade}\left(\dfrac{r_i}{d},\gamma,\nu\right)\right), & \text{point charge model}\\[2ex] \dfrac{U\cdot d_{kj}}{2}\left(h_{jk}^{Pade}\left(\dfrac{r_o}{f\,R_0},\gamma,\nu\right)-2h_{jk}^{Pade}\left(\dfrac{r_i}{f\,R_0},\gamma,\nu\right)\right), & \text{sphere}-\text{plane model}\end{cases},\qquad(33)$$

while $u_1=u_2=0$ as follows from the symmetry considerations. However, note that that elastic conditions on the tube surfaces may be different (e.g. free), necessitating the use of different Green's function.

Effective piezo-response profile in the centre of nested cylindrical domains vs. outer radius $r_o/d$ is shown in Fig.17 for different ferroelectric materials.

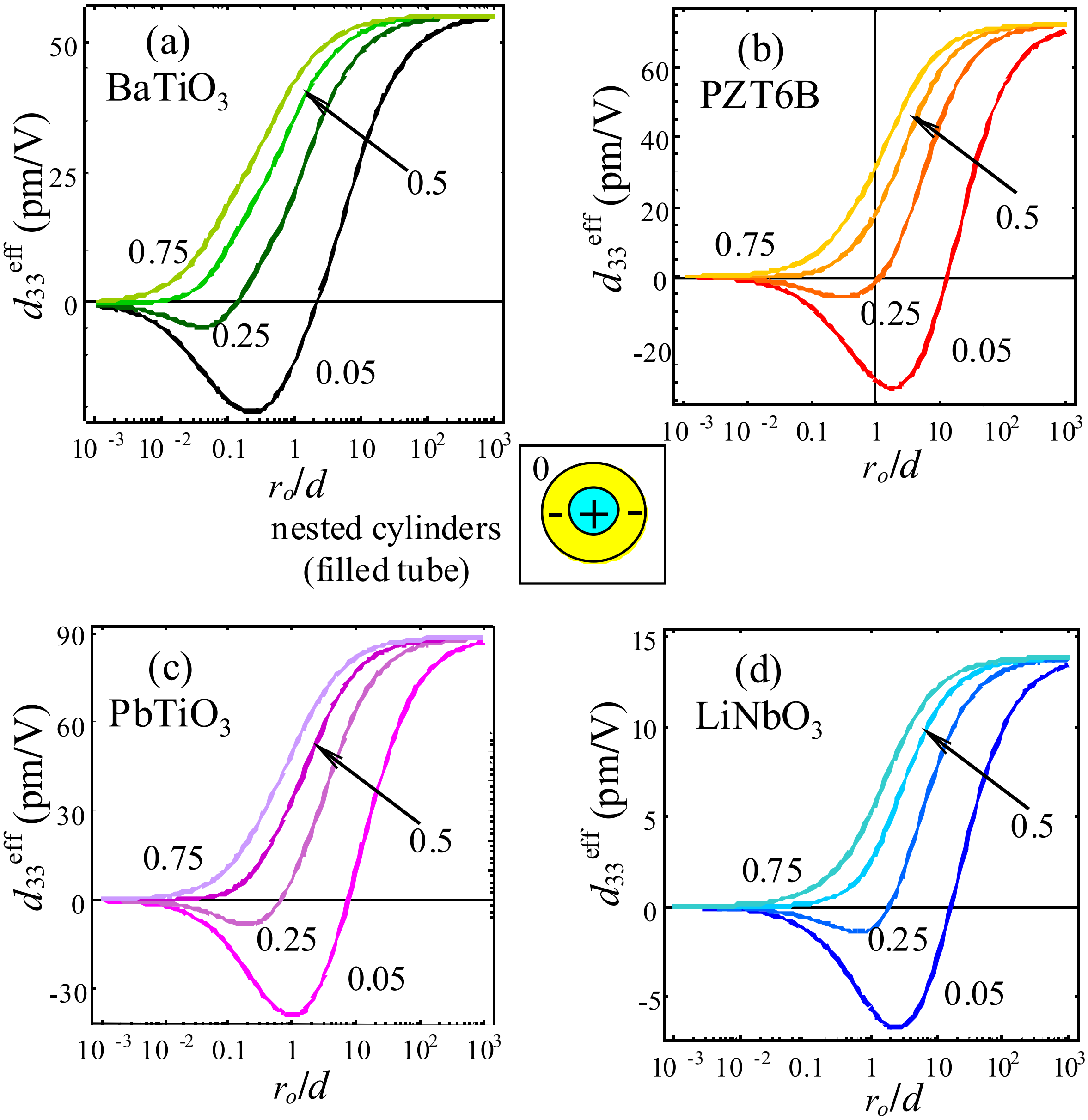

FIG. 17 (Color online). Effective piezo-response profile in the centre of nested cylindrical domain (filled tube) vs. dimensionless outer radius $r_o/d$ for point-charge model of the tip for different ferroelectric materials $BaTiO_3$ (a), PZT6B (b), $PbTiO_3$ (c), $LiNbO_3$ (d). Other

parameters: $\nu = 0.35$; figures near the curves corresponds to the ratio $r_i/r_o = 0.05; 0.25; 0.5; 0.75$.

From the figure, the piezoresponses of a filled domain tube essentially depend on the ratio $r_i/r_o$ (compare different curves). For the tube with $r_i/r_o \geq 0.5$ the response does not change the positive sign, whereas the pronounced negative minimum appeared at $r_o/d \sim 1$ for $r_i/r_o \leq 0.25$. The minimum depth rapidly increases under the ratio $r_i/r_o$ decrease, and ultimately becomes fully saturated to the value corresponding to cylindrical domain.

## VI. Information limit in PFM

The key parameter of any microscopic technique is the minimal size of the object that can be reliable identified. In the context of PFM, this information limit corresponds to the minimal domain size that can be observed experimentally. In the absence of thermal noise on the topographically uniform surface, even infinitely small domain will be visible with signal strength decreasing with size. Practically, this condition is limited by the thermal noise of the cantilever, voltage noise of the bias source, laser noise, etc. Also, the cantilever vibrations induced due to the lateral motion on non-uniform surface will contribute to the effective noise level. The magnitude for some of the noise components can be estimated, e.g. thermal noise in the off resonant conditions is $\langle z^2 \rangle = \sqrt{4k_B TB/(Qk\omega_0)}$, where $k_B$ is Boltzmann constant, $T$ is temperature, $B$ = 1 kHz is the measurement bandwidth, $Q$ is quality factor, $k$ is spring constant, and $\omega_0$ is resonant frequency. Here, we estimate $Q$ = 20 typical for cantilever in contact with the surface, and typical value for $\omega_0$ = 200 kHz. Note that the use of $k$ = 1000

N/m for typical tip-surface contact leads to unphysical small estimates for noise, since the flexural oscillations of the cantilever under distributed thermal loading providing dominant contribution to thermomechanical noise are ignored. Hence, we use $k = 8k_{free}$, where $k_{free}$ is a spring constant of freely vibrating cantilever. The full theory of cantilever dynamic in PFM is given elsewhere.[53]

The estimate of thermomechanical noise yields noise amplitude of ~0.3 pm independently on driving bias. Experimental measurements yield the noise value of ~1 pm/V, close to thermomechanical limit, that will be used here. The maximal contrast $d_{33}^{max}$ in the center of domain stripe(s) and cylinders vs. their sizes (in $d$ units) are shown in Fig.18 for different ferroelectric materials. It is clear that the highest contrast corresponds to a single domain stripe (curves (*s*)), whereas nested tubes (curves (*t*)) reveal the lowest one. Also the single stripe (s) has higher contrast in comparison with their periodic structure (*p*); the piezoresponse in the centre of single cylindrical domain (c) is always higher than the one from nested tubes (t), as it should be expected. Note, that nested cylindrical domains in nonpolar matrix have maximal contrast in their center at $r_i / r_0 \geq 0.5$, that is not valid for domain tubes in polar one (such structure appears during the domain reversal in kinetic limit), where the response could be maximal on the tube (compare Figs. 16 and 18). Again, that the best sensitivity to small domain structures formed below the tip can be achieved in $BaTiO_3$, whereas the worst one corresponds to the $LiNbO_3$.

The cross-over between the signal strength and noise limit yields the information limit of the PFM. Note that virtually in all cases the information limit is well below the resolution, as anticipated. For strongly piezoelectric materials, the domains can be detected even if domain size is well below (by the factor of 100-1000) the characteristic size of the tip. However, for weakly piezoelectric materials the information limit is smaller, as anticipated.

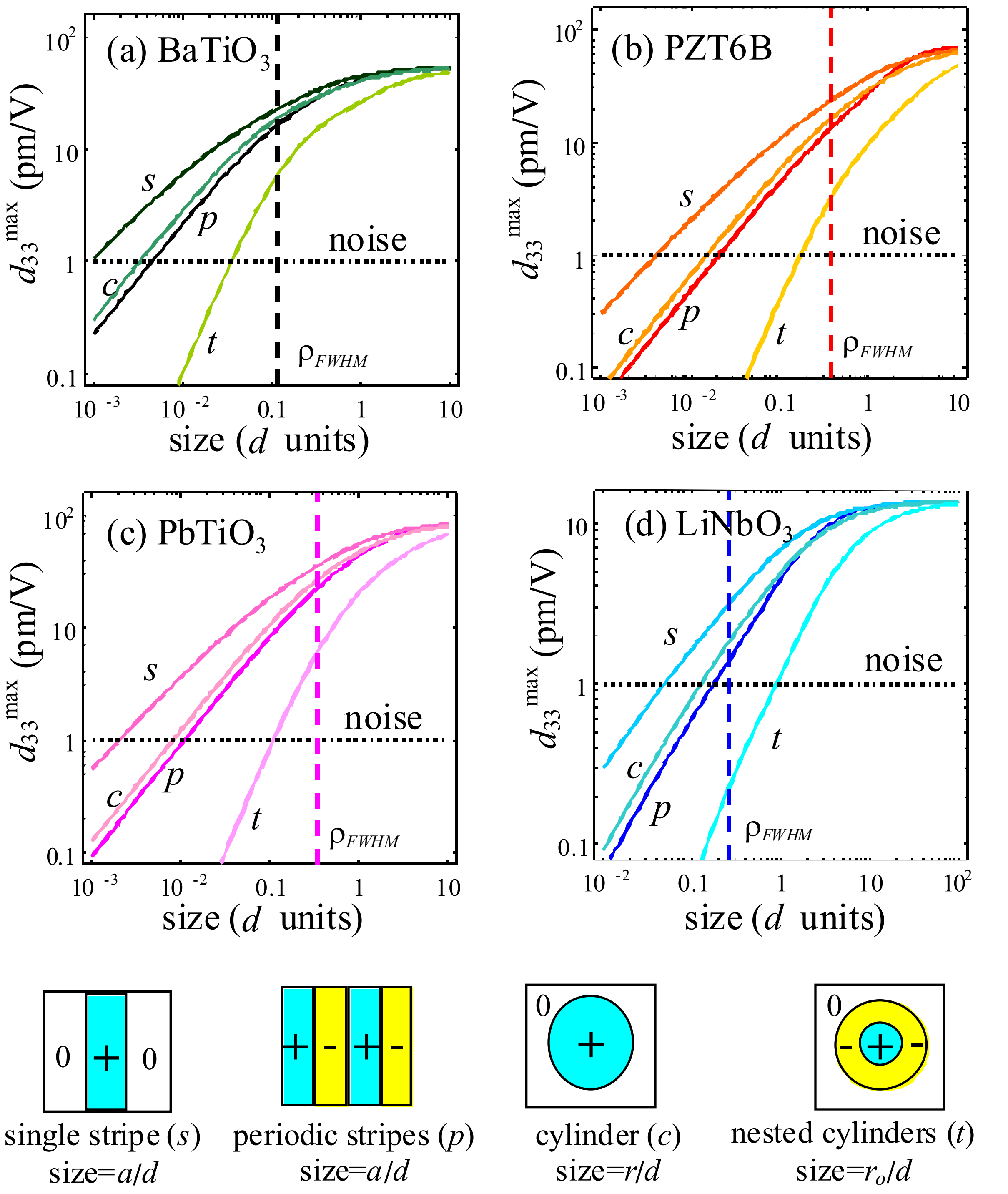

FIG. 18. (Color online). The maximal contrast $d_{33}^{max}$ in the center of domain structures vs. their sizes (in $d$ units) for different ferroelectric materials $BaTiO_3$ (a), PZT6B (b), $PbTiO_3$ (c) and $LiNbO_3$ (d). Letters on the curves correspond to the structures schematically presented below, namely: (*s*) – single domain stripe of thickness $a$ in nonpolar matrix "0", (*p*) – periodic domain stripes of thickness $a$, (*c*) – cylindrical domain of radius $a$ in nonpolar matrix "0", (*t*) – nested cylindrical domains (filled tubes) of outer radius $r_0$ and inner radius $r_i = 0.5 r_0$ in nonpolar matrix "0". Horizontal lines designate the noise level of 1pm/V typical for experiments, whereas vertical ones corresponds to halfwidth $\rho_{FWHM}$ (RTR) calculated from Eq.(12).

## VII. Conclusion

Image formation mechanism in PFM and spectroscopy is analyzed using linear imaging theory. Using Pade analysis of integral representations for PFM signal in decoupled approximation, analytical expressions for resolution and object transfer functions are derived. RTR in PFM is calculated in the point charge and sphere-plane models and effects of dielectric constants and dielectric anisotropy on resolution is determined. The analysis predicts that large transversal dielectric constants and small tip radii favors high-resolution imaging. From this analysis, the highest resolution can be expected for materials such as $BaTiO_3$, while lowest can be expected for materials such as $LiNbO_3$. Predicted RTR is the highest for $BaTiO_3$.

Approximate analytical expressions for vertical and lateral PFM signal vs. distance to domain walls are derived. This potentially allows tip parameters (i.e., image charges distribution representing the tip) to be obtained from PFM signal across the domain wall.

Alternatively, material parameters such as dielectric anisotropy and piezoelectric tensor components can potentially be derived from effective piezoresponse components provided that tip parameters are partially known. The latter is useful for thin ferroelectric films on different substrates, where the material parameters are often thickness and strain dependent and can differ significantly from the bulk ones.

Finally, the analytical form for PFM signal from cylindrical domain is obtained, providing an approach for reconstructing the domain parameters from spectroscopy data.

Research sponsored by the Division of Materials Sciences and Engineering, Office of Basic Energy Sciences, U.S. Department of Energy, under contract DE-AC05-00OR22725 with Oak Ridge National Laboratory, managed and operated by UT-Battelle, LLC.

## APPENDIX A. Fourier representation for Green's function.

Green's function tensor for semiinfinite isotropic elastic half-plane is given by Lurie[41] and Landau and Lifshitz[42]:

$$G_{ij}(x_1,x_2,\xi_3)=\begin{cases}\frac{1+\nu}{2\pi Y}\left[\frac{\delta_{ij}}{R}+\frac{(x_i-\xi_i)(x_j-\xi_j)}{R^3}+\frac{1-2\nu}{R+\xi_3}\left(\delta_{ij}-\frac{(x_i-\xi_i)(x_j-\xi_j)}{R(R+\xi_3)}\right)\right] & i,j\neq 3\\ \frac{(1+\nu)(x_i-\xi_i)}{2\pi Y}\left(\frac{-\xi_3}{R^3}-\frac{(1-2\nu)}{R(R+\xi_3)}\right) & i=1,2\quad and\ j=3\\ \frac{(1+\nu)(x_j-\xi_j)}{2\pi Y}\left(\frac{-\xi_3}{R^3}+\frac{(1-2\nu)}{R(R+\xi_3)}\right) & j=1,2\quad and\ i=3\\ \frac{1+\nu}{2\pi Y}\left(\frac{2(1-\nu)}{R}+\frac{\xi_3^{\ 2}}{R^3}\right) & i=j=3\end{cases}$$

(A.1)

Here $R=\sqrt{(x_1-\xi_1)^2+(x_2-\xi_2)^2+\xi_3^2}$ is radius vector, $Y$ is Young's modulus, and $\nu$ is the Poisson ratio. The components (A.1) depend only on the differences $x_1-\xi_1$ and $x_2-\xi_2$, hence 2D Fourier transform can be introduced as

$$G_{ij}(x_1,x_2,x_3=0,\xi)=\frac{1}{2\pi}\int_{-\infty}^{\infty}dk_1\int_{-\infty}^{\infty}dk_2\,\exp\left(-ik_1(x_1-\xi_1)-ik_2(x_2-\xi_2)\right)\cdot\widetilde{G}_{ij}(k_1,k_2,\xi_3) \qquad \text{(A.2)}$$

where the Green's function components in Fourier representation are:

$$\widetilde{G}_{ij}(k_1,k_2,\xi)=\frac{1+\nu}{2\pi Y}\frac{\exp(-\xi k)}{k}\left(2\delta_{ij}-\frac{k_ik_j}{k^2}(\xi k+2\nu)\right),\quad i,j\neq 3 \qquad \text{(A.3a)}$$

$$\widetilde{G}_{i3}(k_1,k_2,\xi)=-\frac{1+\nu}{2\pi Y}\cdot\frac{ik_i\exp(-\xi k)}{k^2}\left(\xi k+(1-2\nu)\right),\quad i=1,2 \qquad \text{(A.3b)}$$

$$\widetilde{G}_{3j}(k_1,k_2,\xi)=-\frac{1+\nu}{2\pi Y}\cdot\frac{ik_j\exp(-\xi k)}{k^2}\left(\xi k-(1-2\nu)\right),\quad j=1,2 \qquad \text{(A.3c)}$$

$$\widetilde{G}_{33}(k_1,k_2,\xi)=\frac{1+\nu}{2\pi Y}\frac{\exp(-\xi k)}{k}\left(2(1-\nu)+\xi k\right) \qquad \text{(A.3d)}$$

and $k \equiv \sqrt{k_1^2 + k_2^2}$ . The Fourier representation $\widetilde{G}_{ij,l}(k_1, k_2, \xi)$ of the Green's function gradient $\partial G_{ij} / \partial \xi_l$ can be found from Eq. (A.3) as:

$$\widetilde{G}_{ij,l}(k_1, k_2, \xi) \equiv \begin{cases} ik_l \widetilde{G}_{ij}(k_1, k_2, \xi), & l = 1,2 \\ \dfrac{\partial}{\partial \xi} \widetilde{G}_{ij}(k_1, k_2, \xi), & l = 3 \end{cases} \tag{A.4}$$

For the case of dielectrically transverse isotropic ferroelectric the electrostatic potential distribution in the Fourier space is given by

$$\widetilde{V}_Q(k_1, k_2, \xi_3) = \frac{2Q \exp(-kd - k\xi_3/\gamma)}{4\pi\varepsilon_0 k(\varepsilon_e + \kappa)} \tag{A.5}$$

From Eq. (A.5), the electric field $E_k(\mathbf{x}) = -\partial V_Q(\mathbf{x}) / \partial x_k$ . In Fourier space $\widetilde{E}_k(k_x, k_y, x_3)$ can be found as

$$\widetilde{E}_i(k_1, k_2, x_3) \equiv \begin{cases} ik_i \widetilde{V}(k_1, k_2, x_3), & i = 1,2 \\ -\dfrac{\partial}{\partial x_3} \widetilde{V}(k_1, k_2, x_3), & i = 3 \end{cases} \tag{A.6}$$

Resolution function components $\widetilde{W}_{3ij}(q)$ are:

$$\widetilde{W}_{333}(q) \approx -\frac{Q \cdot d}{2\pi\varepsilon_0(\kappa + \varepsilon_e)} \int_0^\infty k dk \int_0^{2\pi} d\psi \exp(-kd) \left( \begin{aligned} &\frac{1}{\gamma\sqrt{k^2 + q^2 - 2kq\cos\psi} + k} + \\ &+ \frac{\gamma\sqrt{k^2 + q^2 - 2kq\cos\psi}}{\left(\gamma\sqrt{k^2 + q^2 - 2kq\cos\psi} + k\right)^2} \end{aligned} \right) \tag{A.7a}$$

$$\widetilde{W}_{313}(q) \approx -\frac{Q \cdot d}{2\pi\varepsilon_0(\kappa + \varepsilon_e)} \int_0^\infty k dk \int_0^{2\pi} d\psi \exp(-kd) \left( \begin{aligned} &\frac{1 + 2\nu}{\gamma\sqrt{k^2 + q^2 - 2kq\cos\psi} + k} - \\ &- \frac{\gamma\sqrt{k^2 + q^2 - 2kq\cos\psi}}{\left(\gamma\sqrt{k^2 + q^2 - 2kq\cos\psi} + k\right)^2} \end{aligned} \right) \tag{A.7b}$$

$$\widetilde{W}_{351}(q) \approx -\frac{Q\cdot d}{2\pi\varepsilon_0(\kappa+\varepsilon_e)}\int_0^\infty k dk \int_0^{2\pi} d\psi \frac{\exp(-kd)}{k} + \frac{\gamma^2\left(k^2 - kq\cos\psi\right)}{\left(\gamma\sqrt{k^2+q^2-2kq\cos\psi}+k\right)^2} \qquad \text{(A.7c)}$$

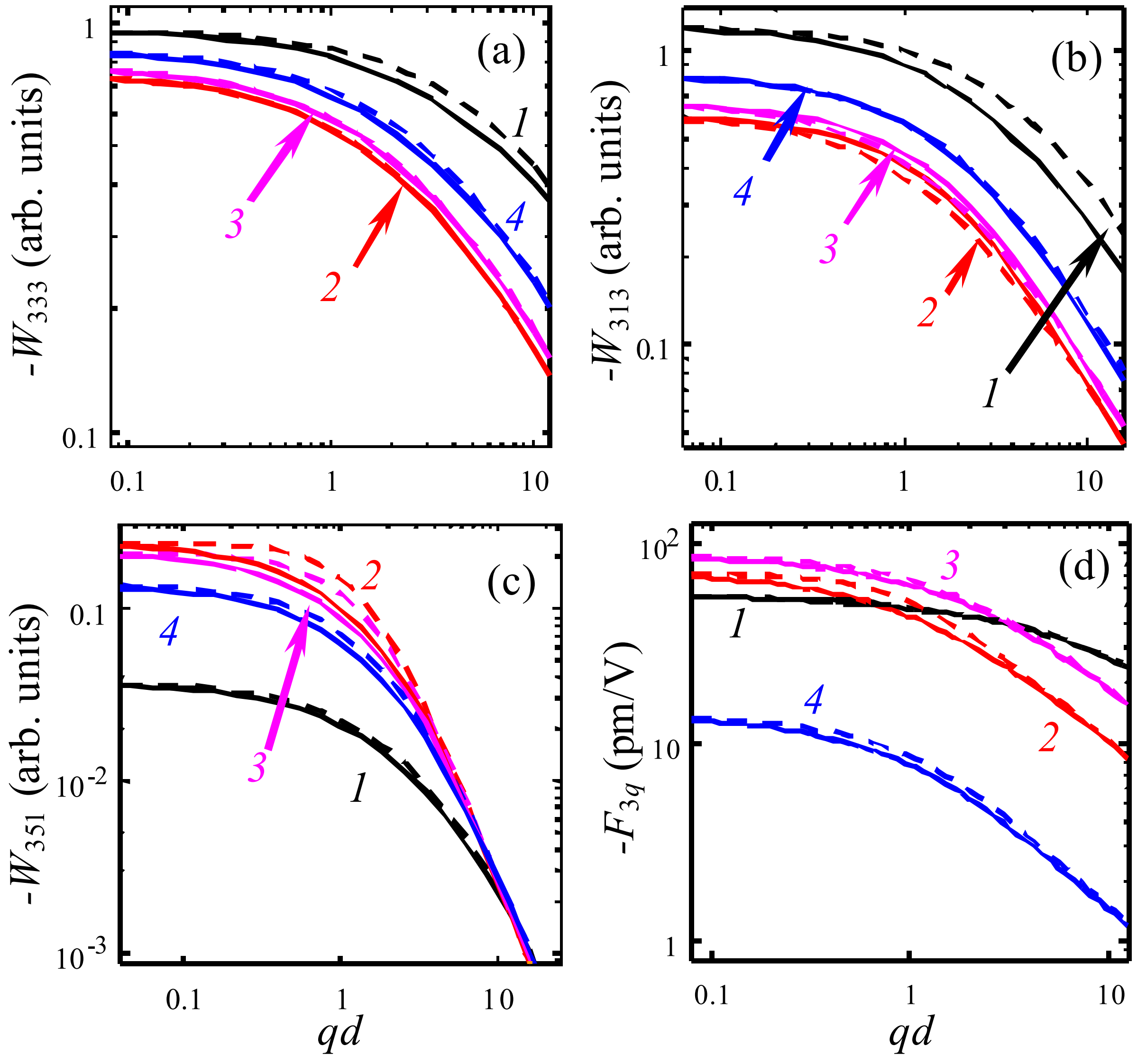

FIG. 1A. (Color online). The resolution function component $\widetilde{W}_{3ij}/V_Q(\mathbf{0})$ (a,b,c) and $F_{3q}(q)$ (d) vs. the wave vector absolute value $q$ for different ferroelectric materials $BaTiO_3$ ( $\gamma = 0.24$, curves 1), PZT6B ($\gamma = 0.99$, curves 2), $PbTiO_3$ ($\gamma = 0.87$, curves 3), $LiNbO_3$ ($\gamma = 0.60$, curves 4). Solid curves - exact expression (A.7), dashed curves - approximation (11b).

Pade approximations for the $\tilde{W}_{333}(q)$ halfwidth:

$$q_{FWHM}=\frac{-1-4\gamma-2\gamma^2+4\gamma^3+(1+2\gamma)\sqrt{17+68\gamma+96\gamma^2+56\gamma^3+20\gamma^4}}{2\gamma\, d(1+2\gamma)^2} \quad \text{(A.8a)}$$

for a point charge model, whereas

$$q_{FWHM}=\frac{\left(\begin{array}{l}(1+\gamma)^2\varepsilon_e+\left(-2-4\gamma+\gamma^2\right)\kappa+\\+\sqrt{5(1+\gamma)^4\varepsilon_e^2+2(1+\gamma)^2\left(2+4\gamma+5\gamma^2\right)\varepsilon_e\kappa+\left(8+32\gamma+36\gamma^2+8\gamma^3+5\gamma^4\right)\kappa^2}\end{array}\right)}{2(1+2\gamma)\gamma\varepsilon_e R_0}$$

(A.8b)

for the sphere-plane one.

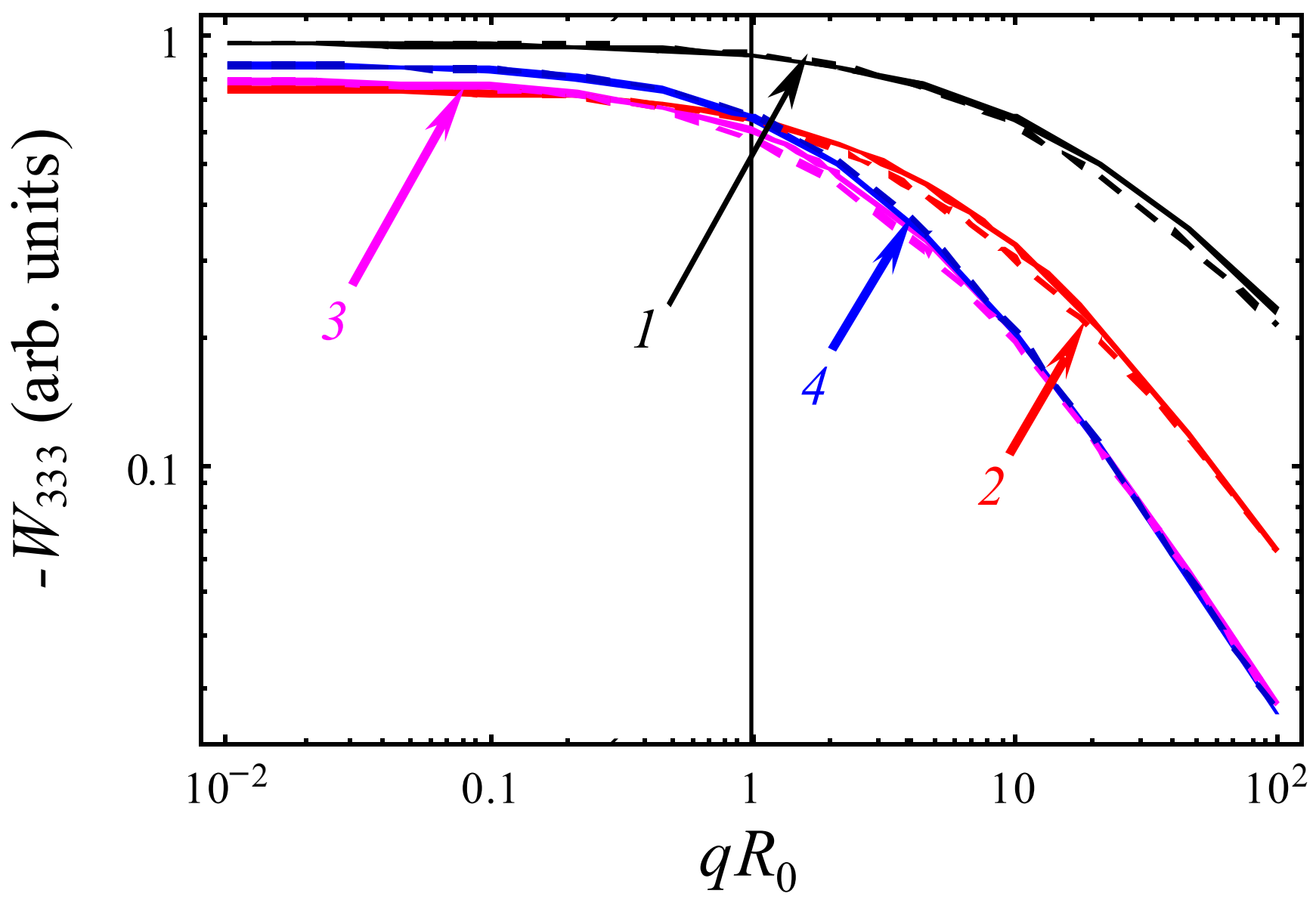

FIG. 2A. (Color online). The resolution function components $\tilde{W}_{333}$ vs. the wave vector absolute value $q$ for different ferroelectric materials $BaTiO_3$ (1), PZT6B (2), $PbTiO_3$ (3), $LiNbO_3$ (4). Calculation was performed within rigorous sphere-plane model of the tip.

Comparison of exact expression (solid curves) with approximation (13a) (dashed curves) is presented.

For the case of dielectrically transverse isotropic ferroelectric under electrostatic potential (3) approximate analytical expressions for the "point charge" resolution function $W_{333}(\rho)$ have been derived, namely Pade:

$$W_{333}(\rho)=\frac{Q}{4\pi^2\varepsilon_0(\kappa+\varepsilon_e)}\begin{cases}-\frac{1}{d^5\gamma^4\rho}\left(2\left(\gamma d\left(d^2\gamma^2+3\gamma d\rho-6\rho^2\right)+3\rho\left(d^2\gamma^2-2\rho^2\right)\ln\left(\frac{\rho}{\gamma d+\rho}\right)\right)\right) \\ \qquad \rho \ll d \\ -\frac{1}{\rho}\cdot\left(\sqrt{\frac{d^2\gamma}{2}}+\frac{4\sqrt{2}\gamma\rho}{\sqrt{3\pi\cdot{}_2F_1\left(3/2,5/2;4;1-1/\gamma^2\right)}}\right)^{-2}, \quad \rho>d\end{cases}$$

(A.9)

Hereinafter ${}_2F_1(p,q;r;s)$ is the hypergeometric function. The resolution function component $W_{333}/V_Q(\mathbf{0})$ vs. the distance $\rho$ for different transversally isotropic ferroelectric materials is depicted in Fig.2A. It is clear from the figure that rather cumbersome Pade approximations are quite precise at $\rho\, d < 0.1$.

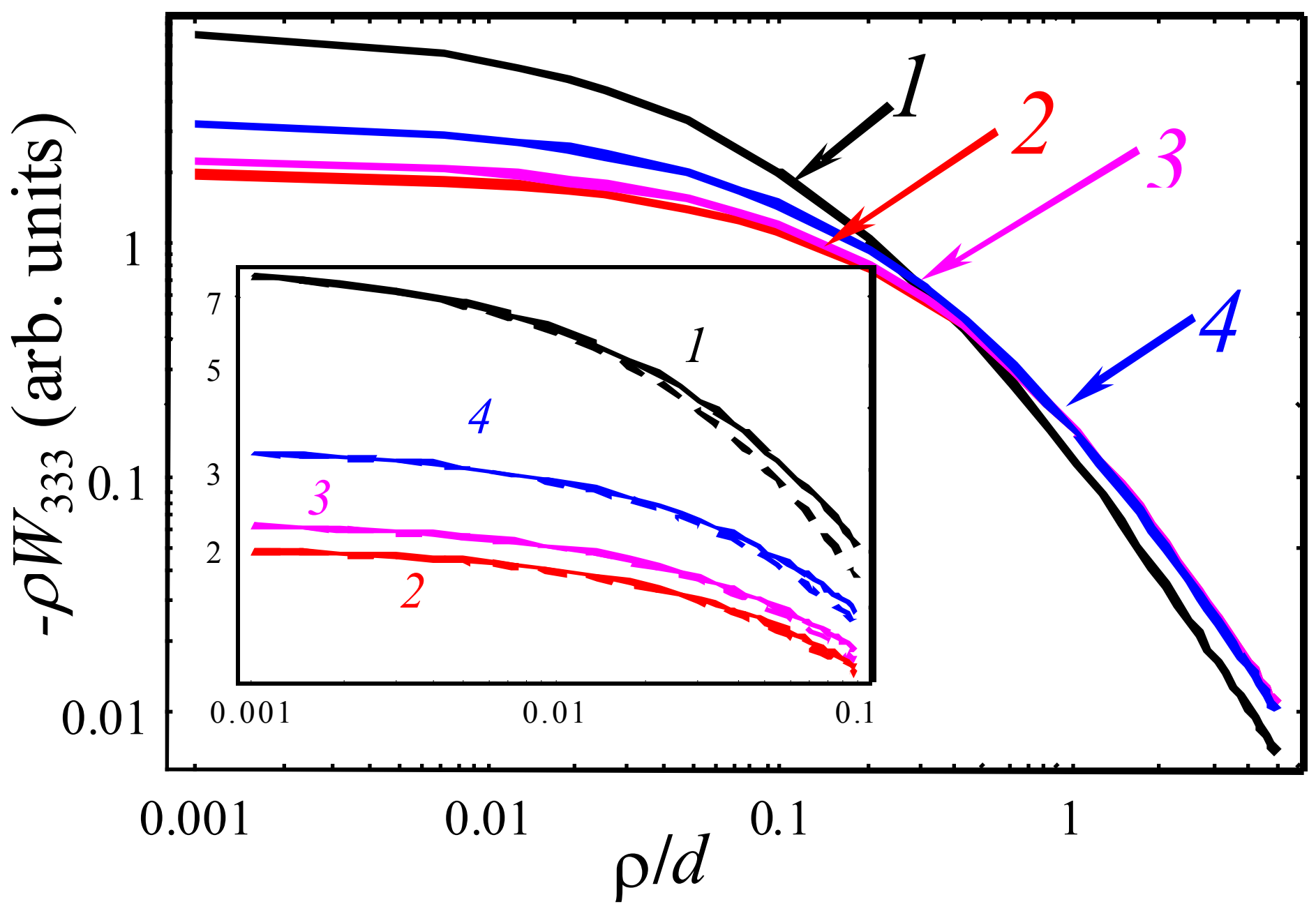

FIG. 3A. (Color online). The "point charge" resolution function component $\rho W_{333}/V_Q(\mathbf{0})$ vs. the distance $\rho$ for different ferroelectric materials $BaTiO_3$ ($\gamma = 0.24$, curves 1), PZT6B ($\gamma = 0.99$, curves 2), $PbTiO_3$ ($\gamma = 0.87$, curves 3), $LiNbO_3$ ($\gamma = 0.60$, curves 4). Comparison of exact expression (8) (solid curves) with approximation (A.7) (dashed curves) is shown on the inset.

## APPENDIX B. Flat domain wall profile.

The potential induced by the point charge above the dielectrically transversely isotropic half plane is:

$$V_Q(\rho,z)=\frac{Q}{2\pi\varepsilon_0(\kappa+\varepsilon_e)}\frac{1}{\sqrt{\rho^2+(z/\gamma+d)^2}},\qquad\text{(B.1)}$$

Here $\sqrt{x^2+y^2}=\rho$ and $z$ are the radial and vertical coordinates respectively, $\kappa=\sqrt{\varepsilon_{33}\varepsilon_{11}}$ is effective dielectric constant of material, $\gamma=\sqrt{\varepsilon_{33}/\varepsilon_{11}}$ is the dielectric anisotropy factor. From Eq. (B.1), the piezo-response in the point (0,0,0) from the infinitely thin domain wall located at distance $a$ from the origin (0,0,0) in the point charge approximation is given by

$$u_i(\mathbf{0},a)=\int_0^\infty dz'\int_{-\infty}^{\infty}dy'\int_{-\infty}^{a}dx'F_i(x',y',z')-\int_0^\infty dz'\int_{-\infty}^{\infty}dy'\int_a^\infty dx'F_i(x',y',z')\qquad\text{(B.2a)}$$

$$F_i(x',y',z')=\begin{pmatrix}e_{311}(G_{i1,1}E_3+G_{i2,2}E_3)+e_{333}G_{i3,3}E_3+\\+e_{113}(G_{i3,1}E_1+G_{i1,3}E_1+G_{i3,2}E_2+G_{i2,3}E_2)\end{pmatrix}\qquad\text{(B.2b)}$$

Using the elastic stiffnesses for isotropic body, the relationship $e_{klj}=d_{kmn}c_{mnlj}$ can be rewritten as follows:

$$\frac{1+\nu}{Y}e_{31}=\frac{d_{31}+\nu d_{33}}{1-2\nu},\quad\frac{1+\nu}{Y}e_{33}=\frac{2\nu d_{31}+(1-\nu)d_{33}}{1-2\nu},\quad\frac{1+\nu}{Y}e_{15}=\frac{d_{15}}{2}\qquad\text{(B.3)}$$

The transition $\hat{e}\to\hat{d}$ employed here noticeably improves the accuracy of the further Pade approximations and conforms to the physics of the problem, as discussed in text. Below, we derive Pade approximations for three components of surface displacement profile across the domain wall.

### B.1. Vertical component.

Taking into account that integrand $F_3(x',y',z')$ in Eq. (B.2a) is an even function on *x'* and *y'*, so (for *a*>0)

$$\int_{-\infty}^{a} dx' F_3(x') - \int_{a}^{\infty} dx' F_3(x') \equiv \int_{-\infty}^{-a} dx' F_3(x') + \int_{-a}^{a} dx' F_3(x') - \int_{a}^{\infty} dx' F_3(x') \equiv \int_{-a}^{a} dx' F_3(x') = 2\int_{0}^{a} dx' F_3(x'),$$

and thus:

$$u_3(\mathbf{0},a) = 4\,\mathrm{sign}(a)\int_0^{\infty} dz' \int_0^{\infty} dy' \int_0^{|a|} dx' F_3(x',y',z') \qquad \text{(B.4a)}$$

Hereinafter $a > 0$ and $F_3(x',y',z')$ is a sum of three terms:

$$G_{33,3}E_3 e_{333} = \frac{Q}{2\pi\varepsilon_0(\kappa+1)}\frac{1+\nu}{2\pi Y}\cdot\frac{(z'/\gamma+d)z'}{\gamma\left(\sqrt{x'^2+y'^2+(z'/\gamma+d)^2}\right)^3}\left(-\frac{3z'^2}{R^5}+\frac{2\nu}{R^3}\right)e_{333} \qquad \text{(B.4b)}$$

$$\left(G_{31,1}+G_{32,2}\right)E_3 e_{311} = \frac{Q}{2\pi\varepsilon_0(\kappa+1)}\frac{1+\nu}{2\pi Y}\cdot\frac{(z'/\gamma+d)z'}{\gamma\left(\sqrt{x'^2+y'^2+(z'/\gamma+d)^2}\right)^3}\left(\frac{3z'^2}{R^5}-\frac{2(1-\nu)}{R^3}\right)e_{311} \qquad \text{(B.4c)}$$

$$\left(G_{33,1}E_1+G_{31,3}E_1+G_{33,2}E_2+G_{32,3}E_2\right)e_{113} = \frac{Q}{2\pi\varepsilon_0(\kappa+1)}\frac{1+\nu}{2\pi Y}\cdot\frac{-6z'^2\left(x'^2+y'^2\right)}{\gamma\left(\sqrt{x'^2+y'^2+(z'/\gamma+d)^2}\right)^3 R^5}e_{113} \qquad \text{(B.4d)}$$

where $R=\sqrt{x'^2+y'^2+z'^2}$. Hence, Eq.(B.4a) can be rewritten as:

$$u_3(\mathbf{0},a) = \frac{Q}{2\pi\varepsilon_0(\varepsilon_e+\kappa)}\frac{1}{d}\left(g_{313}(s,\gamma,\nu)d_{31}+g_{351}(s,\gamma)d_{15}+g_{333}(s,\gamma)d_{33}\right) \qquad \text{(B.5)}$$

where $s=a/d$. After transformation to spherical coordinates, the integration on radius $R$ can be done analytically. After transformation,

$$g_{313}(s,\gamma,\nu)=\frac{2}{\pi}\int_0^{\frac{\pi}{2}}\int_0^{\frac{\pi}{2}}\frac{s\left(3\cos^2\theta-2(1+\nu)\right)\sin\theta\cos\theta\, d\theta\, d\varphi}{\gamma\sqrt{\left(\sin\theta\cos\varphi+\frac{s}{\gamma}\cos\theta\right)^2+s^2\sin^2\theta}} \tag{B.6a}$$

$$g_{351}(s,\gamma)=-3\frac{2}{\pi}\int_0^{\frac{\pi}{2}}\int_0^{\frac{\pi}{2}}\sin\theta\cos^2\theta\left(1-\frac{\sin\theta\cos\varphi+\frac{s}{\gamma}\cos\theta}{\sqrt{\left(\sin\theta\cos\varphi+\frac{s}{\gamma}\cos\theta\right)^2+s^2\sin^2\theta}}\right)d\theta d\varphi \tag{B.6b}$$

$$g_{333}(s,\gamma)=-3\frac{2}{\pi}\int_0^{\frac{\pi}{2}}\int_0^{\frac{\pi}{2}}\frac{s\cdot\sin\theta\cos^3\theta\, d\theta d\varphi}{\gamma\sqrt{\left(\sin\theta\cos\varphi+\frac{s}{\gamma}\cos\theta\right)^2+s^2\sin^2\theta}} \tag{B.6c}$$

Here, we derive Pade approximations for Eqs. (B.6a-c) and compare them with numerical calculations. For the cases $s >> 1$ and $0 < s << 1$, one can easily find the first two terms of the asymptotic series for these integrals as follows

$$g_{313}(s,\gamma,\nu)=\begin{cases}\frac{1+2\gamma}{(1+\gamma)^2}\left(1-\frac{C_{333}(\gamma)}{s}\right)-2\frac{1+\nu}{1+\gamma}\left(1-\frac{C_{313}(\gamma)}{s}\right), & s>>1\\ -\frac{2}{\pi}\frac{s}{\gamma}\left(1+2\ln\left(\frac{s}{\gamma}\right)\right)+(1+\nu)\frac{4}{\pi}\frac{s}{\gamma}\ln\left(\frac{s}{\gamma}\right), & s<<1\end{cases}, \tag{B.7a}$$

$$g_{351}(s,\gamma)=\begin{cases}-\frac{\gamma^2}{(1+\gamma)^2}\left(1-\frac{C_{351}(\gamma)}{s}\right), & s>>1\\ -\frac{\gamma}{2\pi(1-\gamma^2)^{3/2}}\left(2\sqrt{1-\gamma^2}-\gamma^2\ln\left(\frac{1+\sqrt{1-\gamma^2}}{1-\sqrt{1-\gamma^2}}\right)\right)s, & s<<1\end{cases} \tag{B.7b}$$

$$g_{333}(s,\gamma)=\begin{cases}-\frac{1+2\gamma}{(1+\gamma)^2}\left(1-\frac{C_{333}(\gamma)}{s}\right), & s>>1\\ \frac{2}{\pi}\frac{s}{\gamma}\left(1+2\ln\left(\frac{s}{\gamma}\right)\right), & s<<1\end{cases} \tag{B.7c}$$

Here $C_{313}(\gamma)=\frac{1+\gamma}{8\gamma^2}\,{}_2F_1\left(\frac{3}{2},\frac{3}{2};3;1-\frac{1}{\gamma^2}\right)$, $C_{333}(\gamma)=\frac{3(1+\gamma)^2}{16\gamma^2(1+2\gamma)}\,{}_2F_1\left(\frac{3}{2},\frac{5}{2};4;1-\frac{1}{\gamma^2}\right)$,

$C_{351}(\gamma)=\frac{3(1+\gamma)^2}{16\gamma^2}\,{}_2F_1\left(\frac{3}{2},\frac{3}{2};4;1-\frac{1}{\gamma^2}\right)$ and ${}_2F_1(p,q;r;s)$ is the hypergeometric function, in particular $C_{313}(1)=\frac{1}{4}$, $C_{333}(1)=\frac{1}{4}$, $C_{351}(1)=\frac{3}{4}$. Taking into account that $u_3(\mathbf{0},a=0)=0$ and using Eqs.(B.7), two point Pade approximations[54] exponentially tailored with expressions for small $s$ values were found for expressions (B.6), namely:

$$\begin{aligned} g_{351}^{Pade}(s,\gamma)\approx&-\frac{\gamma^2}{2\pi(1-\gamma^2)^{3/2}}\left(2\sqrt{1-\gamma^2}-\gamma^2\ln\left(\frac{1+\sqrt{1-\gamma^2}}{1-\sqrt{1-\gamma^2}}\right)\right)\frac{s}{\gamma}\cdot\exp\left(-\mu\frac{|s|}{\gamma}\right)-\\ &-\frac{\gamma^2}{(1+\gamma)^2}\cdot\frac{s}{|s|+C_{351}(\gamma)}\left(1-\exp\left(-\mu\frac{|s|}{\gamma}\right)\right)\end{aligned}, \quad \text{(B.8a)}$$

$$g_{333}^{Pade}(s,\gamma)\approx\frac{2}{\pi}\frac{s}{\gamma}\left(1+2\ln\left(\frac{|s|}{\gamma}\right)\right)\exp\left(-\mu\frac{|s|}{\gamma}\right)-\frac{1+2\gamma}{(1+\gamma)^2}\cdot\frac{s}{|s|+C_{333}(\gamma)}\left(1-\exp\left(-\mu\frac{|s|}{\gamma}\right)\right),$$

(B.8b)

$$\begin{aligned} g_{313}^{Pade}(s,\gamma,\nu)\approx&\frac{s}{\gamma}\left(-\frac{2}{\pi}\left(1+2\ln\left(\frac{|s|}{\gamma}\right)\right)+(1+\nu)\frac{4}{\pi}\ln\left(\frac{|s|}{\gamma}\right)\right)\exp\left(-\mu\frac{|s|}{\gamma}\right)+\\ &+\left(\frac{1+2\gamma}{(1+\gamma)^2}\cdot\frac{s}{|s|+C_{333}(\gamma)}-2\frac{1+\nu}{1+\gamma}\cdot\frac{s}{|s|+C_{313}(\gamma)}\right)\left(1-\exp\left(-\mu\frac{|s|}{\gamma}\right)\right)\end{aligned}. \quad \text{(B.8c)}$$

Here $\mu$ is the fitting parameter that weakly depends on $\gamma$. Namely, $\mu\approx 5-10$ in the region $\gamma\in[0.1...10]$. Comparison of Eqs. (B.6) - (B.8) are depicted in Fig.1B. Note, that exact curves (B.6) almost coincide with Pade-exponential tailoring (B.8) at $\mu\approx 5-10$, whereas simple Pade approximations (i.e., $\mu\to\infty$) rigorously speaking are invalid at $s\to 0$ (compare solid and dashed curves in Fig. 1B).

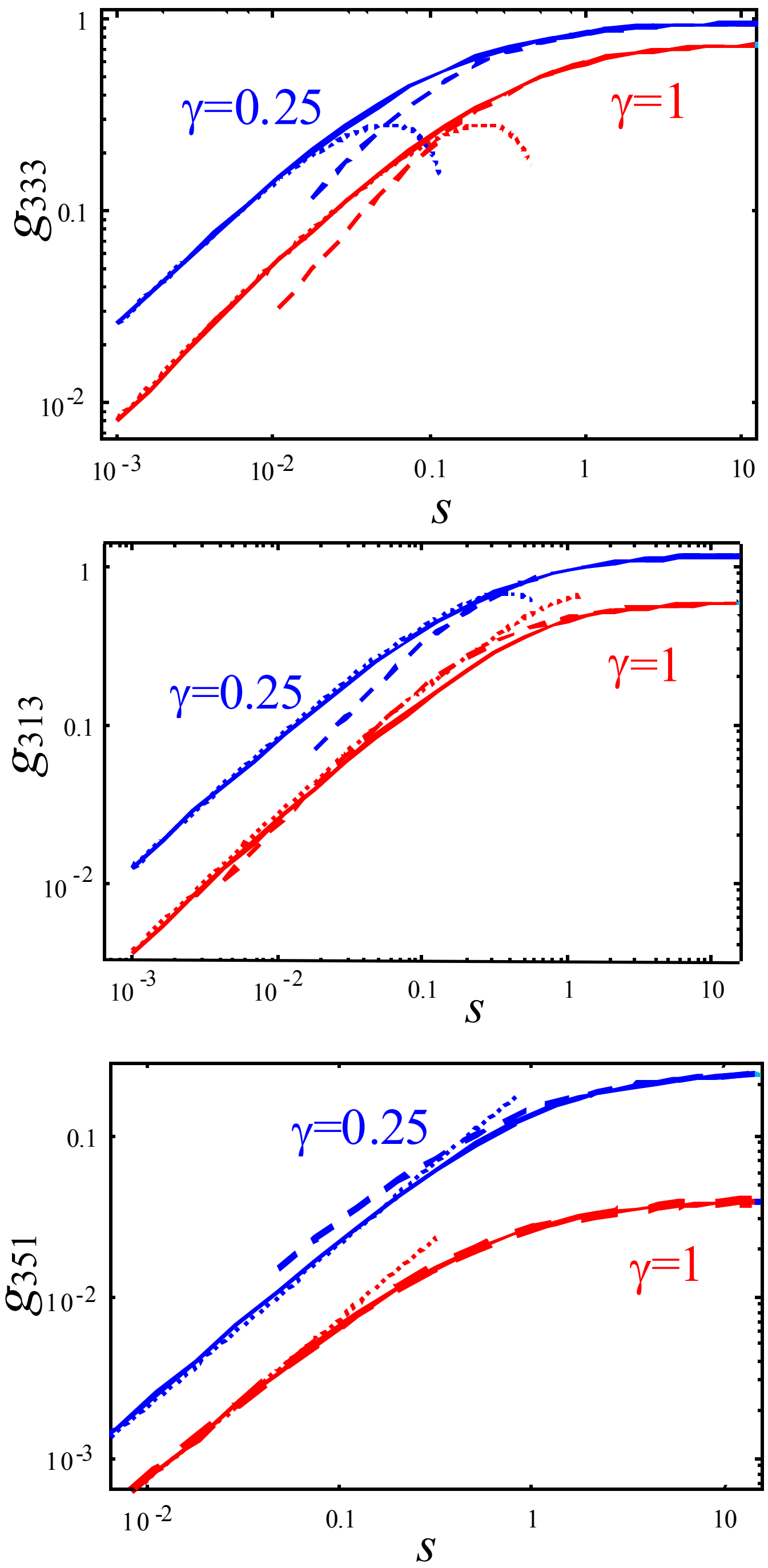

FIG. 1B. (Color online). Comparison of Eqs. (B.6) (solid curves) with expansions (B.7) at small s (dotted curves) and Pade approximations (B.8) at $\mu \to \infty$ (dashed curves) for $\nu = 0.35$, $\gamma = 0.25$ and $\gamma = 1$. Note, that exact curves (B.13) almost coincide with Pade-exponential tailoring (B.15) at $\mu \approx 5 - 10$.

In the framework of the sphere–plane model displacement (B.2)-(B.4) are summarized over the image charges $Q_m$ located at distances $d_m$ above the surfaces. This procedure is very cumbersome and even for the Pade approximations (B.8) could be done only numerically. However displacement (B.5) with the asymptotic series (B.7) can be easily summed on $Q_m$ and $d_m$. Namely, using two obvious identities $\sum_{m=0}^{\infty}\frac{Q_m}{d_m}=2\pi\varepsilon_0(\varepsilon_e+\kappa)U$ (voltage on the surface below the tip) and $\sum_{m=0}^{\infty}Q_m=2\pi\varepsilon_0(\varepsilon_e+\kappa)UR_0\frac{2\varepsilon_e}{\kappa-\varepsilon_e}\ln\left(\frac{\varepsilon_e+\kappa}{2\varepsilon_e}\right)$ (tip capacitance), Eq. (B.5) can be rewritten as follows:

$$u_3(\mathbf{0},a>>R_0)=U\left(\begin{array}{l}\left(\frac{1+2\gamma}{(1+\gamma)^2}\left(1-C_{333}(\gamma)\frac{R_0}{a}f\right)-2\frac{1+\nu}{1+\gamma}\left(1-C_{313}(\gamma)\frac{R_0}{a}f\right)\right)d_{31}+\\+-\frac{\gamma^2}{(1+\gamma)^2}\left(1-C_{351}(\gamma)\frac{R_0}{a}f\right)d_{15}+-\frac{1+2\gamma}{(1+\gamma)^2}\left(1-C_{333}(\gamma)\frac{R_0}{a}f\right)d_{33}\end{array}\right)$$

where $f=\frac{2\varepsilon_e}{\kappa-\varepsilon_e}\ln\left(\frac{\varepsilon_e+\kappa}{2\varepsilon_e}\right)$. Using the method above, we derive Pade approximations for $u_3(\mathbf{0},a)$ valid in the entire region of $a$, namely $s\leftrightarrow a/R_0$:

$$g_{351}(s,\gamma)\approx-\frac{\gamma^2}{(1+\gamma)^2}\frac{s}{s+C_{351}(\gamma)f}, \tag{B.9a}$$

$$g_{333}(s,\gamma)\approx-\frac{1+2\gamma}{(1+\gamma)^2}\frac{s}{s+C_{333}(\gamma)f}, \tag{B.9b}$$

$$g_{313}(s,\gamma,\nu)\approx\frac{1+2\gamma}{(1+\gamma)^2}\frac{s}{s+C_{333}(\gamma)}-2\frac{1+\nu}{1+\gamma}\frac{s}{s+C_{313}(\gamma)f}. \tag{B.9c}$$

Similar rigorous Pade-exponential tailoring like (B.8) is unavailable for the sphere-plane model, necessitating summation procedure at $s\to 0$.

## B.2. In plane component perpendicular to the domain wall.

Taking into account that integrand $F_1(x',y',z')$ in Eqs. (B.2) is an odd function of $x'$ and an even function of $y'$, for $a$>0 we obtain

$$\int_{-\infty}^{a} dx' F_1(x') - \int_{a}^{\infty} dx' F_1(x') = \int_{-\infty}^{-a} dx' F_1(x') + \int_{-a}^{a} dx' F_1(x') - \int_{a}^{\infty} dx' F_1(x') = -2\int_{a}^{\infty} dx' F_1(x').$$

For arbitrary $a$:

$$u_1(\mathbf{0},a) = -4\int_0^{\infty} dz' \int_0^{\infty} dy' \int_{|a|}^{\infty} dx' F_1(x',y',z') \tag{B.10}$$

Function $F_1(x',y',z')$ is a sum of three terms:

$$G_{13,3}E_3 = \frac{Q}{2\pi\varepsilon_0(\kappa+1)}\frac{1+\nu}{2\pi Y}\cdot\frac{(z'/\gamma+d)x'}{\gamma\left(\sqrt{x'^2+y'^2+(z'/\gamma+d)^2}\right)^3}\left(-\frac{3z'^2}{R^5}+\frac{2\nu}{R^3}\right) \tag{B.11a}$$

$$\left(G_{11,1}+G_{12,2}\right)E_3 = \frac{Q}{2\pi\varepsilon_0(\kappa+1)}\frac{1+\nu}{2\pi Y}\cdot\frac{(z'/\gamma+d)x'}{\gamma\left(\sqrt{x'^2+y'^2+(z'/\gamma+d)^2}\right)^3}\left(\frac{3z'^2}{R^5}-\frac{2(1-\nu)}{R^3}\right) \tag{B.11b}$$

$$\left(G_{13,1}E_1+G_{11,3}E_1+G_{13,2}E_2+G_{12,3}E_2\right) = \frac{Q}{2\pi\varepsilon_0(\kappa+1)}\frac{1+\nu}{2\pi Y}\cdot\frac{-6z'x'\left(x'^2+y'^2\right)}{\gamma\left(\sqrt{x'^2+y'^2+(z'/\gamma+d)^2}\right)^3 R^5} \tag{B.11c}$$

where $R=\sqrt{x'^2+y'^2+z'^2}$. Eq. (B.10) can then be rewritten as

$$u_1(\mathbf{0},a) = \frac{Q}{2\pi\varepsilon_0(\varepsilon_e+\kappa)}\frac{1}{d}\left(g_{113}(s,\gamma,\nu)d_{31}+g_{151}(s,\gamma)d_{15}+g_{133}(s,\gamma)d_{33}\right) \tag{B.12}$$

where $s=a/d$. After switching to spherical coordinate system, the integration on radius $R$ can be done analytically. This yields the following expression

$$g_{113}(s,\gamma,\nu) = V_{13}(0,\gamma) + \frac{2}{\pi}\int_0^{\frac{\pi}{2}}\int_0^{\frac{\pi}{2}} \frac{s\left(3\cos^2\theta - 2(1+\nu)\right)\sin^2\theta\cos\varphi\, d\theta\, d\varphi}{\gamma\sqrt{\left(\sin\theta\cos\varphi + \frac{s}{\gamma}\cos\theta\right)^2 + s^2\sin^2\theta}}, \qquad \text{(B.13a)}$$

$$g_{151}(s,\gamma) = V_{51}(0,\gamma) - 3\frac{2}{\pi}\int_0^{\frac{\pi}{2}}\int_0^{\frac{\pi}{2}} \sin^2\theta\cos\theta\cos\varphi\left(1 - \frac{\sin\theta\cos\varphi + \frac{s}{\gamma}\cos\theta}{\sqrt{\left(\sin\theta\cos\varphi + \frac{s}{\gamma}\cos\theta\right)^2 + s^2\sin^2\theta}}\right)d\theta\, d\varphi,$$

(B.13b)

$$g_{133}(s,\gamma) = \frac{2}{\pi}\int_0^{\frac{\pi}{2}}\int_0^{\frac{\pi}{2}} d\theta\, d\varphi \frac{3}{\gamma}\left(\frac{\sin^2\theta\cos^2\theta\cos\varphi}{\sqrt{\left(\frac{\cos\theta}{\gamma}\right)^2 + \sin^2\theta}} - \frac{s\cdot\sin^2\theta\cos^2\theta\cos\varphi}{\sqrt{\left(\sin\theta\cos\varphi + \frac{s}{\gamma}\cos\theta\right)^2 + s^2\sin^2\theta}}\right).$$

(B.13c)

Where: $V_{13}(0,\gamma) = -\frac{3}{8\gamma}\,{}_2F_1\left(\frac{1}{2},\frac{3}{2};3;1-\frac{1}{\gamma^2}\right) + \frac{1+\nu}{\gamma}\,{}_2F_1\left(\frac{1}{2},\frac{1}{2};2;1-\frac{1}{\gamma^2}\right),$

$V_{51}(0,\gamma) = \frac{2}{\pi} - \frac{3}{8\gamma}\,{}_2F_1\left(\frac{1}{2},\frac{3}{2};3;1-\frac{1}{\gamma^2}\right)$, $V_{33}(0,\gamma) = \frac{3}{8\gamma}\,{}_2F_1\left(\frac{1}{2},\frac{3}{2};3;1-\frac{1}{\gamma^2}\right).$

Below, we derive Pade approximations and compare them with numerical calculations. For the cases $s >> 1$ and $0 < s << 1$ integrals (B.13) can be evaluated as

$$g_{113}(s,\gamma,\nu) = \begin{cases} \left(-\frac{\gamma}{(1+\gamma)^3} + \frac{1+\nu}{(1+\gamma)^2}\right)\frac{1}{s}, & s >> 1 \\ -C_{133}(\gamma) + (1+\nu)C_{113}(\gamma) - \frac{(1+2\nu)}{\gamma}s, & s << 1 \end{cases}, \qquad \text{(B.14a)}$$

$$g_{151}(s,\gamma)=\begin{cases}\dfrac{(3+\gamma)\gamma^2}{2(1+\gamma)^3}\dfrac{1}{s}, & s>>1\\[2ex] \dfrac{2}{\pi}-C_{133}(\gamma)-\dfrac{1}{2\pi}\left(2\ln\left(\dfrac{2}{s}\right)+\dfrac{1}{\sqrt{1-\gamma^2}}\ln\left(\dfrac{1+\sqrt{1-\gamma^2}}{1-\sqrt{1-\gamma^2}}\right)\right)s^2, & s<<1\end{cases} \quad \text{(B.14b)}$$

$$g_{133}(s,\gamma)=\begin{cases}\dfrac{\gamma}{(1+\gamma)^3}\dfrac{1}{s}, & s>>1\\[2ex] C_{133}(\gamma)-\dfrac{s}{\gamma}, & s<<1\end{cases} \quad \text{(B.14c)}$$

Here $C_{133}(\gamma)=\frac{3}{8\gamma}\cdot {}_2F_1\left(\frac{1}{2},\frac{3}{2};3;1-\frac{1}{\gamma^2}\right)$, in particular $C_{133}(1)=\frac{3}{8}$; $C_{113}(\gamma)=\frac{1}{\gamma}\cdot {}_2F_1\left(\frac{1}{2},\frac{1}{2};2;1-\frac{1}{\gamma^2}\right)$, in particular $C_{113}(1)=1$. Here ${}_2F_1(p,q;r;s)$ is the hypergeometric function. Pade-exponential approximations have the form:

$$g_{133}(s,\gamma)\approx\left(C_{133}(\gamma)-\frac{|s|}{\gamma}\right)\exp\left(-\mu\frac{|s|}{\gamma}\right)+\frac{C_{133}(\gamma)}{1+\frac{(1+\gamma)^3}{\gamma}C_{133}(\gamma)|s|}\left(1-\exp\left(-\mu\frac{|s|}{\gamma}\right)\right), \quad \text{(B.15a)}$$

$$g_{151}(s,\gamma)\approx\left(\frac{2}{\pi}-C_{133}(\gamma)-\frac{1}{2\pi}\left(2\ln\left(\frac{2}{|s|}\right)+\frac{1}{\sqrt{1-\gamma^2}}\ln\left(\frac{1+\sqrt{1-\gamma^2}}{1-\sqrt{1-\gamma^2}}\right)\right)s^2\right)\exp\left(-\mu\frac{|s|}{\gamma}\right)+$$
$$+\frac{1}{\frac{1}{2/\pi-C_{133}(\gamma)}+\frac{2(1+\gamma)^3}{(3+\gamma)\gamma^2}|s|}\left(1-\exp\left(-\mu\frac{|s|}{\gamma}\right)\right), \quad \text{(B.15b)}$$

$$g_{113}(s,\gamma,\nu)\approx\left(-C_{133}(\gamma)+(1+\nu)C_{113}(\gamma)-\frac{(1+2\nu)}{\gamma}|s|\right)\exp\left(-\mu\frac{|s|}{\gamma}\right)-$$
$$-\left(\frac{C_{133}(\gamma)}{1+\frac{(1+\gamma)^3}{\gamma}C_{133}(\gamma)|s|}-\frac{(1+\nu)C_{113}(\gamma)}{1+\frac{(1+\gamma)^2}{\gamma}C_{113}(\gamma)|s|}\right)\left(1-\exp\left(-\mu\frac{|s|}{\gamma}\right)\right). \quad \text{(B.15c)}$$

Here $\mu$ is the fitting parameter that weakly depends on $\gamma$. Namely, $\mu \approx 5-10$ in the region $\gamma \in [0.1...10]$, whereas exact curves (B.13) almost coincide with Pade-exponential (B.15), whereas simple Pade approximations (i.e., $\mu \to \infty$) rigorously speaking are invalid at $s \to 0$ (compare solid and dashed curves in Fig. 2B). Comparison of Eqs. (B.13-15) are depicted in Fig. 2B.

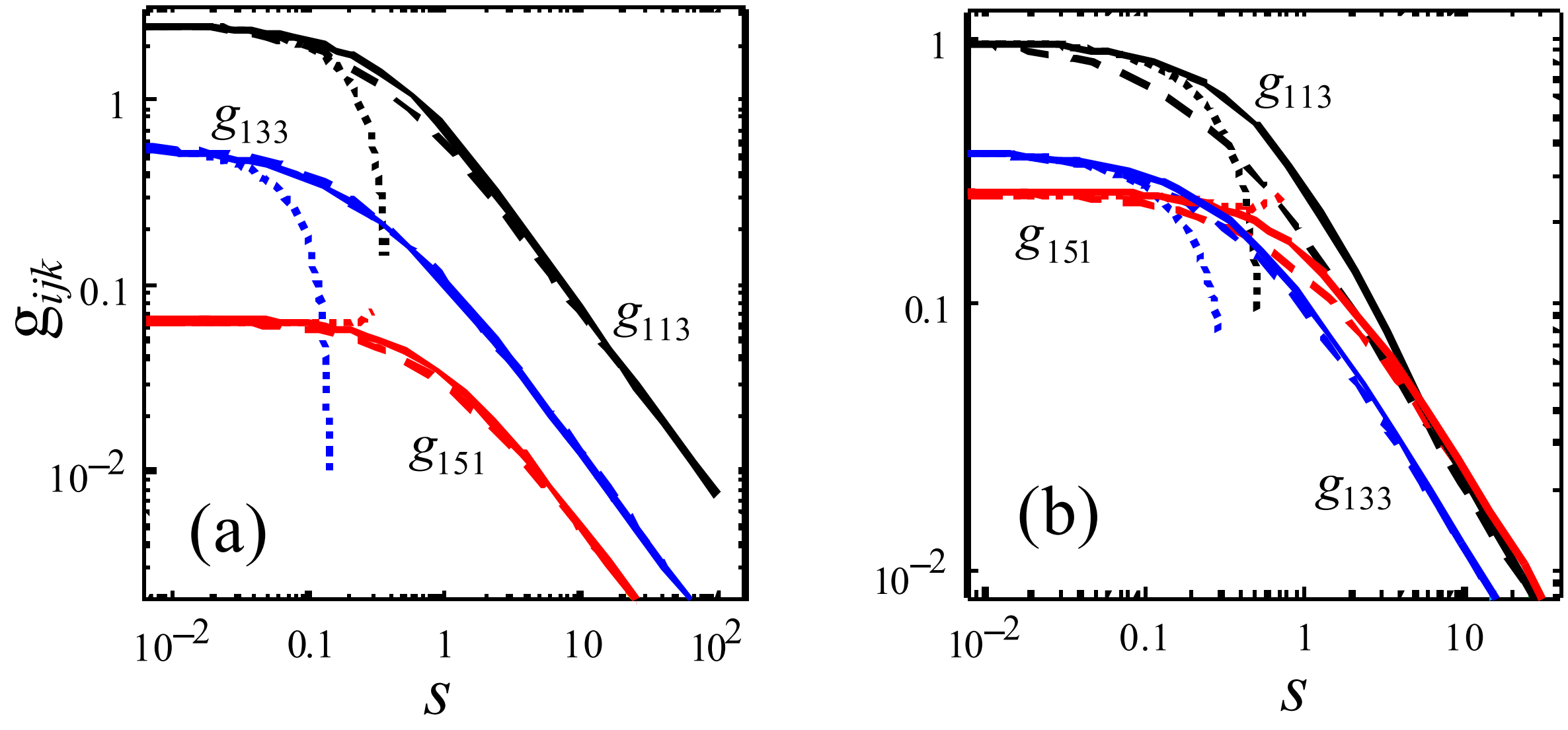

FIG. 2B. (Color online). Comparison of Eqs. (B.13) (solid curves) with (B.14) at small s (dashed and dotted curves) and Pade (B.15) at $\mu \to \infty$ (dashed curves) at $\nu = 0.35$, $\gamma = 0.25$ (a), $\gamma = 1$ (b).

**B.3. In-plane component parallel to the domain wall.**

Since the integrand $F_2(x',y',z')$ in Eqs. (B.2) is an even function on $x'$ and an odd function of on $y'$, $u_2(\mathbf{0}, a) \equiv 0$.

**B.4. Resolution effect on imaging periodic domain structures.**

In Eq. (22) we use that exactly from (21):

$$\widetilde{d}_{klj}(\mathbf{q}) = d_{klj}\delta(q_2)\sum_{n=0}^{\infty}\frac{2i}{(2n+1)\pi}\left(\delta\left(q_1+\frac{2\pi}{a}(2n+1)\right)-\delta\left(q_1-\frac{2\pi}{a}(2n+1)\right)\right) \qquad \text{(B.17)}$$

Thus, in accordance with convolution theorem (9)

$$\widetilde{u}_i(\mathbf{q}) = \widetilde{d}_{klj}(\mathbf{q})\widetilde{W}_{ijkl}(-\mathbf{q}) = \widetilde{W}_{ijkl}(q)d_{klj}\delta(q_2)\sum_{n=0}^{\infty}\frac{2i}{(2n+1)\pi}\left(\delta\left(q_1+\frac{2\pi}{a}(2n+1)\right)-\delta\left(q_1-\frac{2\pi}{a}(2n+1)\right)\right)$$

(B.18)

where $q=\sqrt{q_1^2+q_2^2}$ . The original acquires the form of Eq. (22):

$$u_3(\mathbf{y}) = \int_{-\infty}^{\infty}dq_1\int_{-\infty}^{\infty}dq_2\,\exp(-iq_1y_1-iq_2y_2)\widetilde{u}_3(\mathbf{q}) = \int_{-\infty}^{\infty}dq_1\int_{-\infty}^{\infty}dq_2\,\exp(-iq_1y_1-iq_2y_2)\widetilde{W}_{3jkl}(q)\times$$

$$\times\, d_{klj}\delta(q_2)\sum_{n=0}^{\infty}\frac{2i}{(2n+1)\pi}\left(\delta\left(q_1+\frac{2\pi}{a}(2n+1)\right)-\delta\left(q_1-\frac{2\pi}{a}(2n+1)\right)\right) =$$

$$= \int_{-\infty}^{\infty}dq_1\,\exp(-iq_1y_1)\widetilde{W}_{3jkl}(|q_1|)d_{klj}\sum_{n=0}^{\infty}\frac{2i}{(2n+1)\pi}\left(\delta\left(q_1+\frac{2\pi}{a}(2n+1)\right)-\delta\left(q_1-\frac{2\pi}{a}(2n+1)\right)\right) =$$

$$= d_{klj}\sum_{n=0}^{\infty}\frac{4\widetilde{W}_{3jkl}\left(\frac{2\pi}{a}(2n+1)\right)}{(2n+1)\pi}\,\frac{\exp\left(i\frac{2\pi}{a}(2n+1)\right)-\exp\left(-i\frac{2\pi}{a}(2n+1)\right)}{2i} =$$

$$= \sum_{n=0}^{\infty}\frac{4d_{klj}}{(2n+1)\pi}\widetilde{W}_{3jkl}\left(\frac{2\pi}{a}(2n+1)\right)\sin\left(\frac{2\pi}{a}(2n+1)y_1\right) = \sum_{n=0}^{\infty}\frac{4}{(2n+1)\pi}\sin\left(\frac{2\pi}{a}(2n+1)y_1\right)F_q\left(\frac{2\pi}{a}(2n+1)\right)$$

## APPENDIX C. Cylindrical domain response.

The piezoresponse in the point (0,0,0) from the infinitely thin cylindrical domain wall located at radius $r$ from the origin (0,0,0) in the point charge approximation is given by

$$u_3(\mathbf{0},r) = 2\pi\left(\int_0^\infty dz'\int_r^\infty \rho d\rho' F(\rho',z') - \int_0^\infty dz'\int_0^r \rho d\rho' F(\rho',z')\right)$$
$$F(\rho',z') = \begin{pmatrix} e_{311}(G_{31,1}E_3 + G_{32,2}E_3) + e_{333}G_{33,3}E_3 + \\ + e_{113}(G_{33,1}E_1 + G_{31,3}E_1 + G_{33,2}E_2 + G_{32,3}E_2) \end{pmatrix} \qquad \text{(C.1)}$$

Where $G_{ij,k}$ components are given by Eqs. (A.1). Eq. (C.1) could be rewritten as

$$u_3(\mathbf{0},r) = 2\pi\left(\int_0^\infty dz'\int_0^\infty \rho d\rho' F(\rho',z') - 2\int_0^\infty dz'\int_0^r \rho d\rho' F(\rho',z')\right) \qquad \text{(C.2)}$$

Using the elastic stiffnesses for isotropic body, the relationship (A.4), we can rewrite Eq. (C.2) in the form

$$u_3(\mathbf{0},r) = \frac{Q}{2\pi\varepsilon_0(\varepsilon_e+\kappa)}\frac{1}{d}\left(h_{13}(s,\gamma,\nu)d_{31} + h_{51}(s,\gamma)d_{15} + h_{33}(s,\gamma)d_{33}\right) \qquad \text{(C.3)}$$

where $s = r/d$. After switching to spherical coordinate system the integration on radius $R$ can be done analytically. This yields the following expression

$$h_{13}(s,\gamma,\nu) = \frac{1+2\gamma}{(1+\gamma)^2} - \frac{2(1+\nu)}{1+\gamma} - 2\int_0^{\frac{\pi}{2}} \frac{s\left(3\cos^2\theta - 2(1+\nu)\right)\sin\theta\cos\theta\, d\theta}{\gamma\sqrt{\left(\sin\theta + \frac{s}{\gamma}\cos\theta\right)^2 + s^2\sin^2\theta}} \qquad \text{(C.4a)}$$

$$h_{51}(s,\gamma) = -\frac{\gamma^2}{(1+\gamma)^2} + 6\int_0^{\frac{\pi}{2}} \sin\theta\cos^2\theta\left(1 - \frac{\sin\theta + \frac{s}{\gamma}\cos\theta}{\sqrt{\left(\sin\theta + \frac{s}{\gamma}\cos\theta\right)^2 + s^2\sin^2\theta}}\right)d\theta \qquad \text{(C.4b)}$$

$$h_{33}(s,\gamma)=-\frac{1+2\gamma}{(1+\gamma)^2}+6\int_0^{\frac{\pi}{2}}\frac{s\cdot\sin\theta\cos^3\theta d\theta}{\gamma\sqrt{\left(\sin\theta+\frac{s}{\gamma}\cos\theta\right)^2+s^2\sin^2\theta}} \qquad \text{(C.4c)}$$

For the cases $s>>1$ and $0<s<<1$, these integrals can be approximated by expressions:

$$h_{13}(s,\gamma,\nu)\approx\begin{cases}\frac{1+2\gamma}{(1+\gamma)^2}-\frac{2(1+\nu)}{1+\gamma}-2\left(\frac{1+2\gamma}{(1+\gamma)^2}\left(1-\frac{D_{33}(\gamma)}{s}\right)-\frac{2(1+\nu)}{1+\gamma}\left(1-\frac{D_{13}(\gamma)}{s}\right)\right), & s>>1\\ \frac{1+2\gamma}{(1+\gamma)^2}-\frac{2(1+\nu)}{1+\gamma}+\nu\frac{4s}{\gamma}, & s<<1\end{cases}$$

(C.5a)

$$h_{51}(s,\gamma)\approx\begin{cases}-\frac{\gamma^2}{(1+\gamma)^2}+2\frac{\gamma^2}{(1+\gamma)^2}\left(1-\frac{D_{51}(\gamma)}{s}\right), & s>>1\\ -\frac{\gamma^2}{(1+\gamma)^2}+s^2, & s<<1\end{cases}. \qquad \text{(C.5b)}$$

$$h_{33}(s,\gamma)\approx\begin{cases}-\frac{1+2\gamma}{(1+\gamma)^2}+2\frac{1+2\gamma}{(1+\gamma)^2}\cdot\left(1-\frac{D_{33}(\gamma)}{s}\right), & s>>1\\ -\frac{1+2\gamma}{(1+\gamma)^2}+\frac{4s}{\gamma}, & s<<1\end{cases}, \qquad \text{(C.5c)}$$

Functions $D_{51}(\gamma)=\frac{3\pi(1+\gamma)^2}{32\gamma^4}\left[-2\gamma^2{}_2F_1\left(\frac{1}{2},\frac{3}{2};3;1-\frac{1}{\gamma^2}\right)+{}_2F_1\left(\frac{3}{2},\frac{3}{2};4;1-\frac{1}{\gamma^2}\right)\right]$, in particular $D_{51}(1)=\frac{3\pi}{8}$; $D_{33}(\gamma)=\frac{3\pi(1+\gamma)^2}{32\gamma^2(1+2\gamma)}\cdot{}_2F_1\left(\frac{3}{2},\frac{5}{2};4;1-\frac{1}{\gamma^2}\right)$, in particular $D_{33}(1)=\frac{\pi}{8}$; $D_{13}(\gamma)=\pi\frac{1+\gamma}{16\gamma^2}{}_2F_1\left(\frac{3}{2},\frac{3}{2};3;1-\frac{1}{\gamma^2}\right)$, in particular $D_{13}(1)=\frac{\pi}{8}$. Here ${}_2F_1(p,q;r;s)$ is the hypergeometric function.

Pade-exponential approximations for expressions $h_{jk}^{Pade}(s,\gamma,\nu)$ were found, namely:

$$h_{51}^{Pade}(s,\gamma) \approx -\frac{\gamma^2}{(1+\gamma)^2} + 2\frac{\gamma^2}{(1+\gamma)^2}\cdot\frac{s^2}{2\frac{\gamma^2}{(1+\gamma)^2} + D_{51}(\gamma)s + s^2}, \tag{C.6a}$$

$$h_{33}^{Pade}(s,\gamma) \approx -\frac{1+2\gamma}{(1+\gamma)^2} + 2\frac{1+2\gamma}{(1+\gamma)^2}\cdot\left(\frac{s\cdot\exp\left(-\mu\frac{s}{\gamma}\right)}{s+\frac{\gamma(1+2\gamma)}{2(1+\gamma)^2}} + \frac{s\left(1-\exp\left(-\mu\frac{s}{\gamma}\right)\right)}{s+D_{33}(\gamma)}\right), \tag{C.6b}$$

$$h_{13}^{Pade}(s,\gamma,\nu) \approx \frac{1+2\gamma}{(1+\gamma)^2} - \frac{2(1+\nu)}{1+\gamma} + 4s\left(\frac{(1+\nu)}{(1+\gamma)s+\gamma} - \frac{1}{\frac{2(1+\gamma)^2 s}{1+2\gamma}+\gamma}\right)\exp\left(-\mu\frac{s}{\gamma}\right) -$$
$$-2\left(\frac{1+2\gamma}{(1+\gamma)^2}\cdot\frac{s}{s+D_{33}(\gamma)} - \frac{2(1+\nu)}{1+\gamma}\cdot\frac{s}{s+D_{13}(\gamma)}\right)\left(1-\exp\left(-\mu\frac{s}{\gamma}\right)\right) \tag{C.6c}$$

Here $\mu$ is the fitting parameter that weakly depends on $\gamma$. Namely, $\mu \approx 1-2$ in the region $\gamma \in [0.1...10]$ for the point-charge model, whereas $\mu \to \infty$ for the sphere-plane model. Note, that exact curves almost coincide with Pade-exponential tailoring (C.6) at $\mu \approx 1-2$, whereas simple Pade approximations (i.e., $\mu \to \infty$) are rather good at $s > 0.1$, but rigorously speaking are invalid at $s \to 0$ (see Fig. 1C for details).

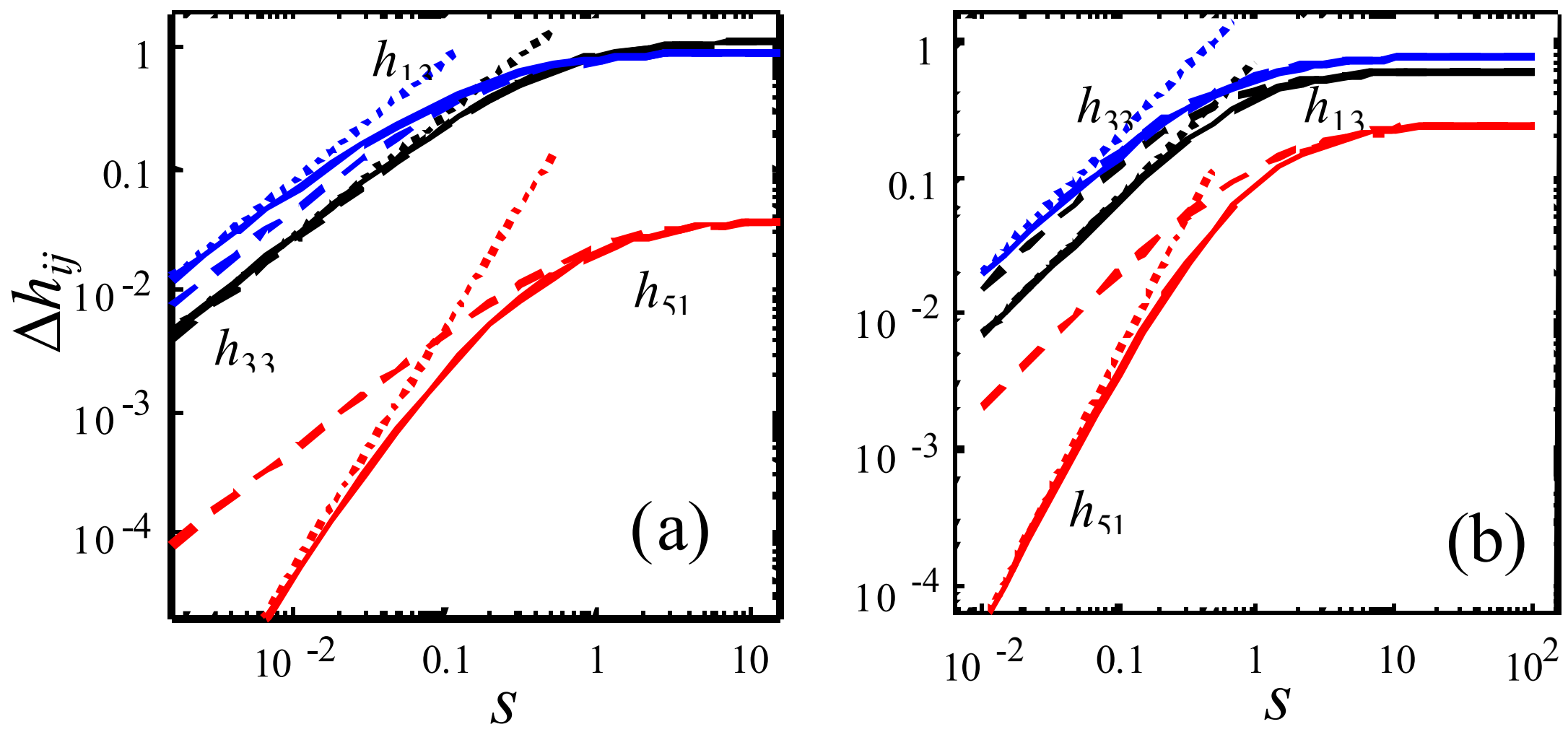

FIG. 1C. (Color online). Comparison of Eqs. (C.4) (solid curves), Pade (C.6) at $\mu \to \infty$ (dashed curves) and (C.5) at small $s$ (dotted curves) for $\Delta h_{ij}(s,\gamma,\nu) = \left(h_{ij}(s,\gamma,\nu) - h_{ij}(0,\gamma,\nu)\right)/2$ at $\nu = 0.35$, $\gamma = 0.25$ (a), $\gamma = 1$ (b). Note, that exact curves (C.4) almost coincide with tailoring (C.6) at $\mu \approx 1-2$.